# Calculating entropy at different scales among diverse communication systems

**GERARDO FEBRES[1], KLAUS JAFFÉ[2]**

[1] *Dept. de Procesos y Sistemas, Universidad Simón Bolívar, Venezuela. mail@gfebres.net*

[2] *Laboratorio de Evolución, Universidad Simón Bolívar, Venezuela. kjaffe@usb.ve*

*We evaluated the impact of changing the observation scale over the entropy measures for text descriptions. MIDI coded Music, computer code and two human natural languages were studied at the scale of characters, words, and at the Fundamental Scale resulting from adjusting the symbols length used to interpret each text-description until it produced minimum entropy. The results show that the Fundamental Scale method is comparable with the use of words when measuring entropy levels in written texts. However, this method can also be used in communication systems lacking words such as music. Measuring symbolic entropy at the fundamental scale allows to calculate quantitatively, relative levels of complexity for different communication systems. The results open novel vision on differences among the structure of the communication systems studied.*

**Key Words:** Entropy; fundamental scale; language recognition; complexity; music entropy; information

## 1. INTRODUCTION

Ever since his influential paper was published in 1948, Shannon's entropy [1] has been the basis for quantifying information of symbolic systems descriptions. In summary, he postulated that the quantity of information of each description is proportional to the description text entropy. Shannon's work was developed over the basis of a binary communication system consisting of two symbols: zeros and ones. But the principle that relates entropy and information applies also to communication systems of more than two symbols.

Quantitative human natural language comparison has been a matter of study for decades. Some studies Schurmann and Grassberger [2] and Kontoyiannis [3], treat their text objects as large sets of characters. Other studies, Savoy [4, 5], Febres, Jaffe and Gershenson [6] and Febres and Jaffe [7], see texts as words. While the study of texts at the scale of characters misses the features of the natural languages that arise due to the structures composed of symbols to form words, the treatment of texts as exclusively built by words, ignores the presence of important elements that are part of the structure of the communication system. Neither the characters nor the words by themselves represent the actual symbol composition of communication systems. The recently proposed concept of *Fundamental Scale* [8], offers the basis for a proper comparison of different types of communication systems. The representation of communication systems at the fundamental scale also allows to include in the comparison communication systems that do not use words.

In this study we compare music, computer programing code, and two natural languages at several observation scales. For each type of communication system we consider a large number of texts. We measure the impact of changing the observation scale over the entropy measured for each text. Quantitative es timates of entropy were obtained at the scale of characters and at the fundamental scale calculated for each communication system [8]. When words are meaningful to the communication system, the scale of words was also included in the comparison.

## 2. METHODS

We measured entropy $h$ and specific diversity $d$ for text descriptions expressed in four different communication systems: English, Spanish, computer programing code and MIDI music. Entropy calculations are used to estimate the quantity of information required for each text description at three different observation scales: characters, words and the fundamental scale.

### 2.1. Diversity and Entropy

A version of Shannon's entropy [1], adapted for communication systems with more than two symbols [4], was used for these calculations. If the communication system $\boldsymbol{B}$ consists of as $D$ different symbols, thus $\boldsymbol{B}$ can be depicted as

$$\boldsymbol{B} = \{Y_1, \dots, Y_i, \dots, Y_D, \ \boldsymbol{P}(Y_i)\} \tag{1}$$

where $Y_i$ represents each symbol, from $Y_1$ to $Y_D$, used in the message, and $\boldsymbol{P}(Y)$ the probability density function which establishes the relative frequencies of appearance of symbols $Y_i$. It is worthwhile to mention that symbols $Y_i$ do not have any syntactical meaning, thus they do not carry any information by themselves.

The quantity of information needed to convey a system description using the set of symbols included in the communication system $\boldsymbol{B}$ can be estimated as its entropy $h$. Thus

$$h = -\sum_{i=1}^{D} p_i \ log_D \ p_i \quad , \tag{2}$$

where $p_i$ refers to the probability of encountering symbol $Y_i$ within a message described using the communication system $\boldsymbol{B}$. Observe that the base of the logarithm is the symbol diversity $D$ and therefore values of entropy $h$ are normalized between zero and one, which is consistent with expressions for normalized entropy values stablished in previous works by Gershenson and Fernandez [5] and Febres and Jaffe [4].

The specific diversity $d$ is a normalized index with values between 0 and 1. It expresses the relationship of the number of symbols used —the diversity $D$— and the description length $N$:

$$d = \frac{D}{N} \quad . \tag{3}$$

A value $d = 1$ means the description uses each symbol exactly once. In this case, and recalling these symbols are strictly symbolic —i.e. they have null syntactic meaning—, no pattern can be formed and thus, to reproduce the description, it would be needed the transcription of the whole set of symbols, employing the maximum quantity of information that possibly fits into a set of $D$ different symbols. Equation (2) produces consistent results, since for this case $p_i = {}^{1}/_{D}$ and therefore, all logarithms within the summation end up being 1 and the entropy reaches its maximum $h = 1$.

The lowest value the diversity can get is $D = 1$. This occurs when the description uses only one symbol and the message consists of a sequence of $N$ identical symbols. When the diversity is $D = 1$, the description can be replaced by indicating the number of symbols $N$, therefore the information needed to express the description is just the number $N$. In this case, $p_1 = 1$, and the summation of Equation (2) will contain only one summand which leads to an undetermined entropy value.

The cases where $D = N$ and $D = 1$ are extremes. In general the diversity $D$ is an integer number between one and $N$. Typically $D$, is larger than 2 and thus, there are $D - 1$ different ways to modify the distribution of symbol probabilities $p_i$, as a result, the entropy $h$ can be thought as a function defined over a dominion of $D - 1$ dimensions.

## 2.2. Language Scale

As signaled in Expression (1), the specific language used in a message can be described as a symbol set along with the associated symbol frequency distribution. But the specific set of symbols considered as part of the language descriptor, depends on the way the whole message is divided in smaller pieces. The criteria used to segment the message into symbols is commonly called the *observation scale*; or simply the *scale*. Therefore an English text, for example, can be interpreted as a set of characters, a set of words, a set of sentences or any other way to rationally organize the information written in pieces. In this study we quantify the term scale as the number of symbols which the whole message is divided, thus the scale is, according to its definition, equal to the language diversity $D$.

We interpret descriptions written in different languages at several scales. We split the messages into characters and words, when the language admits meaningful words. We also interpret the descriptions at their Fundamental Scale as the scale at which the set of symbols lead to a minimal entropy [4].

Music is the superposition of simultaneously performed signals. In contrast to natural languages which can be described as a set of symbols formed by meaningful words, music digital records end up being a sequence of abstract characters. Since the concept of word cannot be applied to music, we do not treat music at the scale of words.

### *2.2.1. The Character's Scale*

To observe a description at the character's scale, the text is segmented as a sequence of single characters. In Expression (1) there will be $D$ different symbols $Y_i$, each of which will be an indivisible character. The random variable $P(Y_i)$ represents the probabilities of occurrence of each character $Y_i$. Since this scale is formed by symbols being of the same size, we classify this scale within the category of *regular size scale*.

### *2.2.2. The Word's Scale*

At the scale of words symbols are made by words or symbols having a comparable function like words within the written text. Words are sequences of characters (different from a space char) preceded and followed by a space or a punctuation sign. There are several considerations to make a precise interpretation of the elements of a text when the scale of words is adopted. Punctuation signs, as considered above, serve as word delimiters. But they also modify the context of the message. Therefore, the punctuation signs have some meaning and should be considered as words themselves. In general, words as symbols are written with lowercase. In English and Spanish the use of capital letters at the beginning of a word indicates it is a proper name or the beginning of an idea just after a period. Still, a word initiating a sentence and written with its first uppercase letter could be a proper name, and thus should be considered a different symbol from that written with the same sequence of letters with all its characters in lower case. An infallible criterion to recognize words, written with subtle variations, as different symbols is nearly impossible. Nonetheless, we built algorithms to recognize most cases where symbol disambiguation is possible. The criteria used for those algorithms is presented in greater detail in our previous study [6]. In our present study we use the same criteria to recognize and classify words as different symbols.

After recognizing all different words existing in a description, language $\boldsymbol{B}$ can be formed by assigning each word to the corresponding instance of variable $Y_i$. The random variable $\boldsymbol{P}(Y_i)$ representing the probabilities of occurrence of each symbol, is determined according to the number of times the symbol $Y_i$ appears in the text and the total number of symbols $N$. Symbols in the scale of words are basically determined by the presence of the space character which works as a symbol delimiter. The symbol lengths is not constant and thus we classify this scale within the category of *symbol delimited-irregular size scale*.

### *2.2.3. The Fundamental Scale*

The Fundamental Scale of a description is a set of symbols that minimizes the description's entropy as expressed in Equation (2). The set of symbols must not have any overlapping as they appear in the description's text. Additionally, when symbols are set one after another at their corresponding places within the text, the exact original description must be reproduced. An algorithm for the determination of the Fundamental Scale in one-dimensional descriptions have been presented by Febres and Jaffe [4]; we rely on it to evaluate the entropy and the complexity of descriptions at this scale. To illustrate the results obtained when analyzing an English text at the Fundamental Scale, we include an example in Appendix A.

## 2.3. Scale Downgrading

The frequency profile associated to a language is a representation of the language. In a language made of $D$ different symbols, this representation uses $D$ values to describe the language. Plotting these values is useful because it permits to graphically observe an abstract description. Depending on the level of detail with which the observer appreciates the description, every value of the frequency of each symbol, may or may not be needed. If for some purpose a rough idea of the profile's shape is sufficient, a smaller number of values can be used. If on the contrary, the observer needs to detail tiny changes in the profile, a higher density of dots will be required to draw these changes of direction. Changing the number of symbols used to describe a system, constitutes a change of the scale of observation of the system; thus we refer to the process of reducing the number of values used to draw the frequency profile as *downgrading the language scale*.

If language $\boldsymbol{B}$ introduced in Equation (1) is employed to build a $N$ symbol long system description, then language $\boldsymbol{B}$ can be specified as the set of $D$ symbols $Y_i$ and the probability density function $\boldsymbol{P}(Y_i)$ which establishes the relative frequencies of appearance of the symbols $f_i$. Thus, using $p_i$ to represent the probability of finding symbol $Y_i$ within the description, we have

$$p_i = p(Y_i) = \frac{f_i}{N} \ , \quad 1 \le i \le D \ . \tag{4}$$

At this point language $\boldsymbol{B}$ is presented at scale $D$. To include the observation scale of a language as part of the nomenclature, we add a sub-index to the letter representing the language. Thus, language $\boldsymbol{B}$ at some scale $S$, where $1 \le S \le D$, would be denoted as $\boldsymbol{B}_S$. When the index does not appear, it can be assumed the language is expressed at its original and maximum scale. That is $\boldsymbol{B} = \boldsymbol{B}_D$.

Notice that in this context, a lower scale signals a smaller number of symbols to depict a communication system. That corresponds to a point of view from which less details —and therefore less symbols— are observed. Figure 1 illustrates how the symbols existing at the original scale $D$ contribute to form groups of symbols which appear at such a probability, that the general shape of the frequency profile is reproduced at the smaller scale $S$.

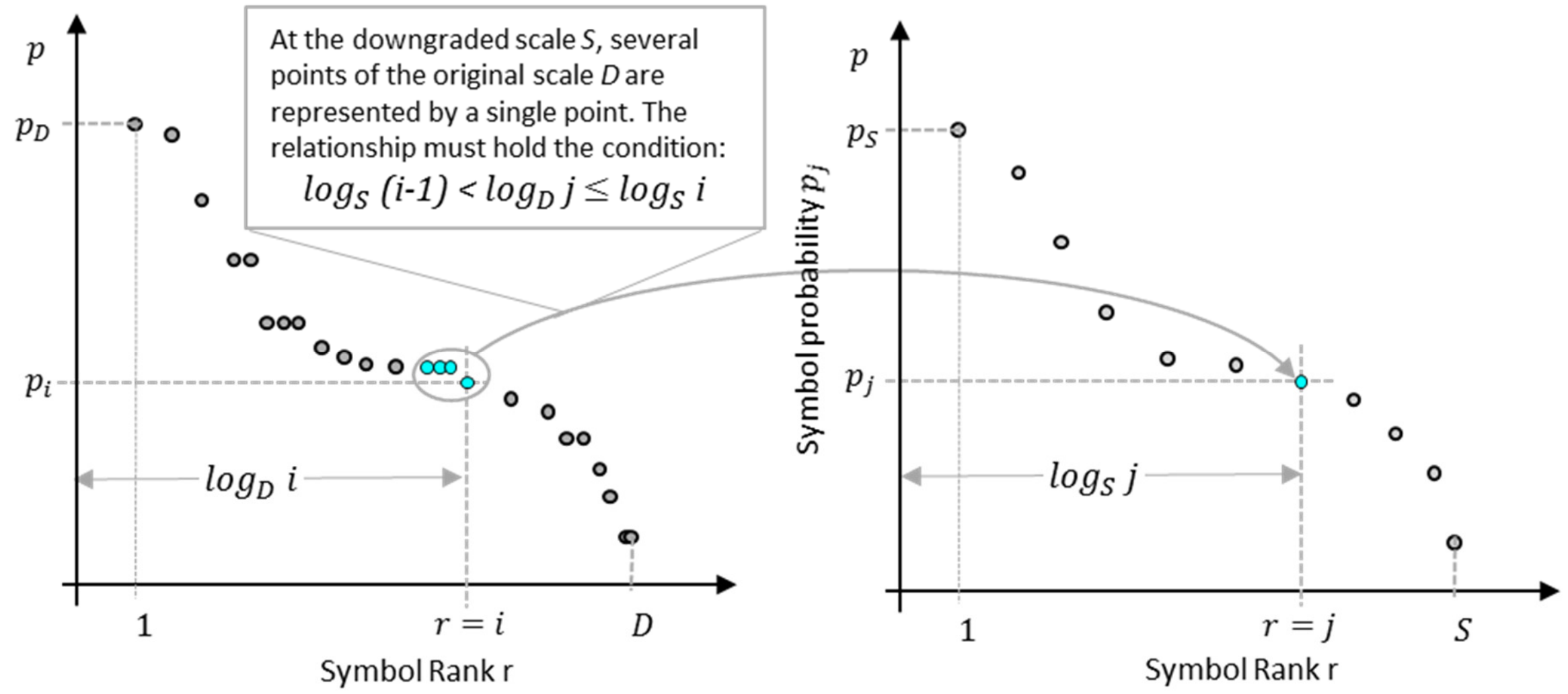

**Figure 1.** Graphic representation of a language scale downgrading from scale $D$ to scale $S$ ($S < D$). The total number of points at scale $D$, representing $D$ symbols on the left graph, are transformed in $S$ points when the language is represented at the scale $S$, as in the right graph.

Downgrading a language from scale $D$ to scale $S$ can be performed by multiplying the transpose of vector $\boldsymbol{P}_D$ by the transformation matrix $\boldsymbol{G}_{D,S}$, as indicated below:

$$\boldsymbol{P}_S = \boldsymbol{P}_D^{\mathrm{T}} \cdot \boldsymbol{G}_{D,S} \,, \tag{5}$$

$$\boldsymbol{G}_{D,S} = \begin{bmatrix} G_{1,1} & G_{1,2} & \cdots & & G_{1,S} \\ G_{2,1} & \ddots & \cdots & \cdots & \vdots \\ \vdots & \vdots & G_{i,j} & & G_{i,S} \\ & \vdots & & \ddots & \vdots \\ G_{D,1} & \cdots & G_{D,j} & \cdots & G_{D,S} \end{bmatrix}, \tag{6a}$$

$$G_{i,j} = \begin{cases} 1 & if \quad log_D(i-1) \leq log_S\, j < log_D\, i \\ 0 & otherwise\,. \end{cases} \qquad 1 \leq i \leq D\,, 1 \leq j \leq S\,, \tag{6b}$$

$$j = int\ (S^{log_D\, i})\,. \tag{6c}$$

This procedure for downgrading the language scale is useful given the frequent requirement of expressing text descriptions at the same scale.

## 2.4. Message Selection.

We applied our methods to messages expressed in three different types of communication systems: natural languages, computer programing code and *MIDI* music. Table 1 shows the number of texts used for each type of communication system.

### *2.4.1. Natural Languages*

As messages written in natural languages we include 128 English and 72 Spanish speeches pronounced by politicians, military, writers, scientists, human right defenders and other public personages. For both, English and Spanish, the length of speeches range from about 200 words to more than 17000 words.

### 2.4.2. Computer Programing Code

Several computer programming codes, obtained from diverse programing languages are included as artificial language descriptions. Comments within the code are usually written in a natural language. Since recognizing these comments is easy, we could clean up most codes and leave them free of natural language comments. Nevertheless, many programing language symbols are created after English and Spanish words. Therefore, avoiding the presence of some natural language words may not be possible. Table 1 shows the different programming communication systems represented in our experiment.

| Meaninful Words | | | |
|---|---|---|---|
| | ***Genre/Class*** | ***Pieces*** | ***Authors*** |
| **English** | Speech | **128** | **108** |
| Sample: I have nothing to offer but blood, toil, tears, and sweat. We have before us an ordeal of the most grievous kind. | | | |
| **Spanish** | Speech | **72** | **56** |
| Sample: Ni en el más delirante de mis sueños, en los días en que escribía Cien Años de Soledad, llegué a imaginar que podría asistir a este acto | | | |
| **Programming Code** | C | **7** | |
| | C Sharp | **20** | |
| | HTML | **2** | |
| | Java | **3** | |
| | Mathlab | **9** | |
| | php | **1** | |
| | Phyton | **1** | |
| | Visual Basic | **4** | |
| Sample: {class Program {void prime_num¡long num¶{ bool isPrime = true; for ¡int i = 0; i = num; i++¶ { for ¡int j = 2; j = num; j++¶ {if ¡i != j && i<j == 0¶ { isPrime = false; | | | |

| Abstract meaning | | | |
|---|---|---|---|
| | ***Period/Style*** | ***Pieces*** | ***Composers*** |
| **MIDI Music** | **Total** | **438** | **> 93** |
| | Medieval | 38 | 12 |
| | Renaissance | 31 | 10 |
| | Baroque | 42 | 8 |
| | Classic | 45 | 7 |
| | Romantic | 89 | 13 |
| | Impressionistic | 34 | 4 |
| | Twenty Century | 35 | 8 |
| | Movie Themes | 18 | > 4 |
| | Rock | 24 | 5 |
| | Hindu Raga | 14 | > 1 |
| | Chinese Traditional | 12 | > 1 |
| | Venezuelan | 56 | > 20 |
| Sample: #dnN Q E / # Nd Qd Ed -d !dn- ! /_ #_nN Q E / # LX OX CX 1Z %Zn1 % 2U &UnL O C 2 & JL NL BL 4O (On4 ( 6J * JnJ N B 6 * E? I? L? @?7E +En7 + 4? (?nE I L @ 4 ( E? J? >? 6? * ?x6?n | | | |

**Table 1.** Number of messages processed for English, Spanish, computer programing code and *MIDI* music. A sample of the texts for each type of communication system is included.

### 2.4.3. MIDI music

Polyphonic music is the result of the superposition of a vast variety of sounds. The information the sheet music may contain a relatively small number of sounds and effects. But the way music sounds, responds not only to the information written on the sheet music. It also brings information about the particular sound of the instrument, the ambient, minor deviations in the pitch and the rhythm, the addition of differences introduced by the interpreter, and innumerable effects, which despite not represented in the music score, are audible and part of the essence of music. Music as written in the sheet music is a discrete information phenomenon, but as it sounds is an analogous process which requires huge information packages to be faithfully recorded.

The digital musical interface *MIDI* is a way of digitizing music as is interpreted. The *MIDI* process converts music into synthetic music. The resulting sequences of discretized sounds are recorded in files with a large, though limited, number of symbols. Taking advantage of this characteristics, we calculated symbol diversity and entropy to hundreds of the almost unlimited *MIDI* music pieces available in Internet.

Most *MIDI* files include metadata at their beginnings and their ends, usually written in English or Spanish. The length of these headers and footers can be considered small compared to the total symbolic description length; since cleaning all files would represent a large non-automated task, we decided not to prune this small amount of noise and leave the files as they show when opened with a *.txt* extension.

## 3. RESULTS

Results are presented in three sets. The first set compares diversity ranges. In a second set the resulting entropy is compared for the communication systems observed at different scales. In the third section we use entropies to calculate the complexity at the fundamental scale for the four types of communication systems considered. In this section we also show an estimation of the length required for messages expressed in each communication system, so that the calculated properties settle down in a stable characteristic value.

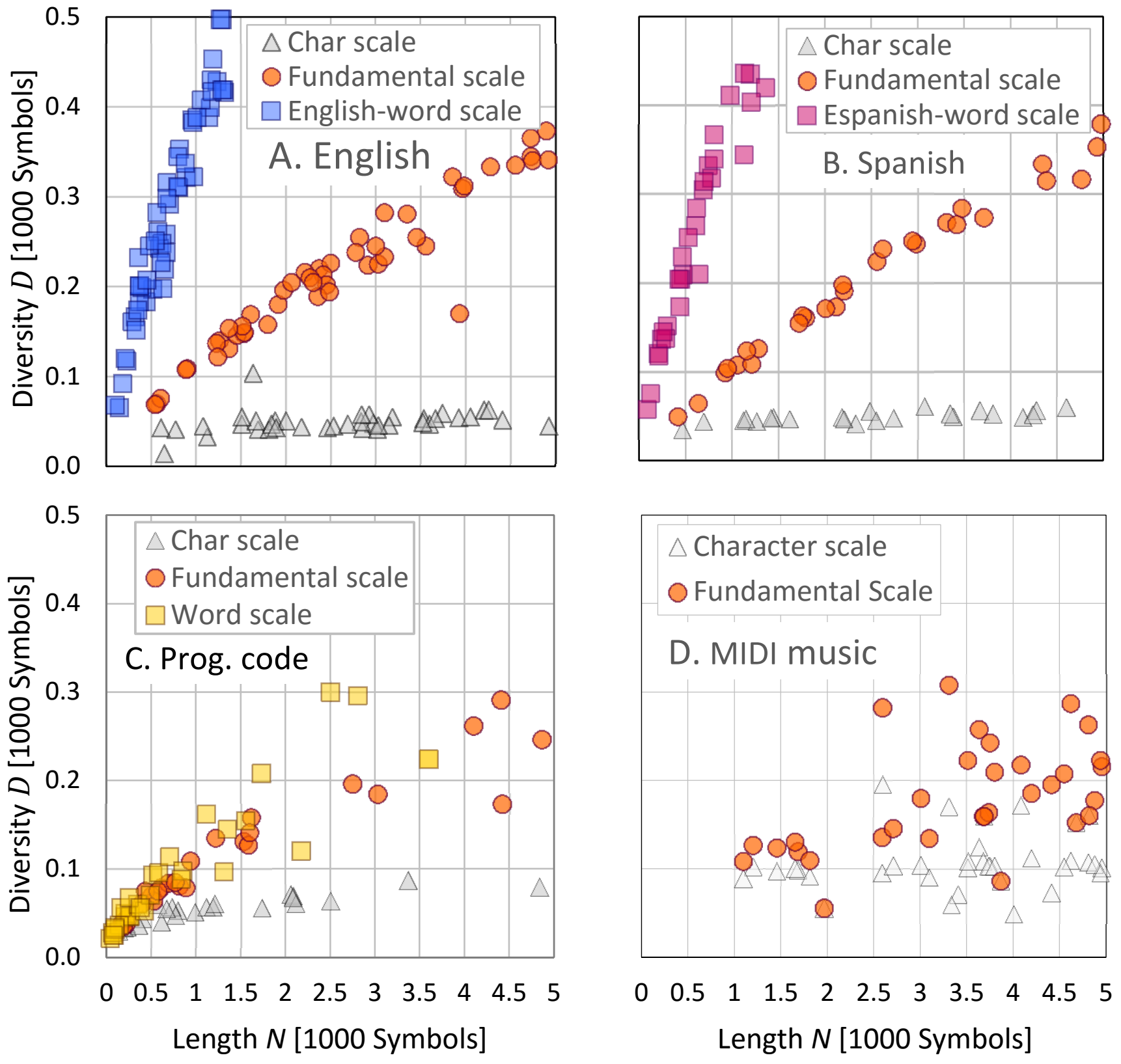

**Figure 2.** Diversity $D$ as a function of the description length $N$ measured in characters. Descriptions expressed in several types of communication systems: A. English, B: Spanish. C: computer programing code and D: *MIDI* music.

## 3.1. Diversity

Figure 2 presents the diversity vs the message length in symbols for the communication systems studied. Independently of the communication system, at the character scale, only different elementary characters may represent symbols. Not being any possibility for combining characters to form strings, as the description length increases, the number of symbols rapidly saturates and cannot grow above certain number on some hundreds.

At the scale of words the graphs show similar results to those exposed in previous studies [6,7]; diversity behaves accordingly to the Heap's law. Since *MIDI* music descriptions do not contemplate meaning for words, therefore is no representation of diversity at this scale for music. At the fundamental scale diversity also follows the Heap's law; as the text length increases the diversity grows, but it grows at a lower speed for longer texts. Something to highlight is the dramatic reduction of diversity dispersion observable at the fundamental scale. Especially for English and Spanish, there seems to be a narrow band of diversity where the symbol diversity should fit in other to achieve a low entropy. For few texts the diversity falls outside this narrow band, but they should be considered as exceptional cases.

## 3.2. Entropy

For each message represented in Table 1, we computed the entropy measured at the scales considered. Figure 3 shows four graphs were entropy is graphed against specific diversity for each communication system. The markers are colored and shaped differently to facilitate the observation of the corresponding scales at which the calculus belong.

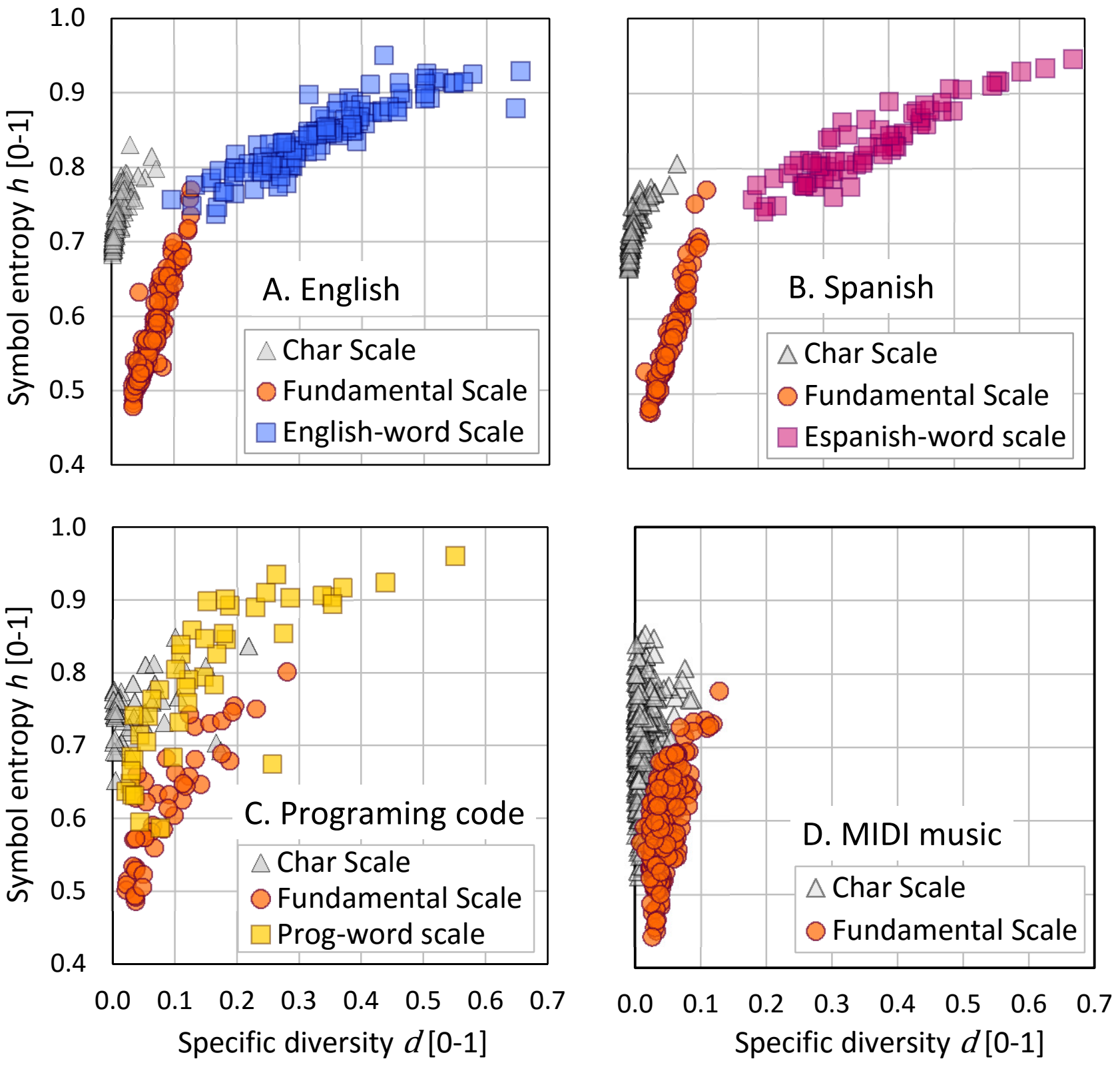

**Figure 3.** Entropy $h$ as a function of specific diversity $d$. Entropies are shown for different observation scales for several types of communication systems: A. English, B: Spanish. C: Computer programing code and D: MIDI music.

As with the diversity, at the fundamental scale the entropy occupies a narrow band in the space entropy-specific diversity. Not surprising here since the symbols were selected precisely to minimize the resulting entropy. For programing code and music the minimal entropy space found is not as reduced as for English and Spanish. Perhaps a consequence of the less restrictive subtypes of communication systems considered as computer programing code and music.

### 3.3. Symbol Frequency Profiles

We compared the profiles of the four communication systems considered. Each communication system is represented as the set of symbols resulting from the union of all the independent pieces belonging to each system. Thus, four large descriptions were processed to produce a profile corresponding to English, Spanish, programming code and MIDI music.

Table 2 shows some properties of these communication systems considered as the union of all descriptions within each class. In order to reduce the number of points from the original number of symbols to 129, the Scale Downgrading calculation explained in Section 2.3 was applied. The Scale Downgrading is useful to normalize the scale of observation bringing the symbol diversity to a specified number while keeping the general shape of the frequency profile shown with log-log axes. Figure 4 shows the resulting symbol frequency profiles representing each communication system studied.

| Communication systems' properties | | | | | | | | |
|---|---|---|---|---|---|---|---|---|
| | Length $N$ | Diversity $D$ | specific diversity $d$ | Entropy $h$ | Approx. transition from single-char. symbol to multiple-char. symbol | | | |
| | | | | | Highest ranked multi-char symbol | | Lowest ranked single-char symbol | |
| | | | | | Symbol | Rank | Symbol | Rank |
| English | 1626927 | 7597 | 0.00467 | 0.430 | of | 34 | Y | 145 |
| Spanish | 984044 | 4688 | 0.00476 | 0.440 | que | 32 | G | 152 |
| Prog. code | 958547 | 2964 | 0.00309 | 0.541 | maxChild | 52 | Y | 247 |
| MIDI music | 14592192 | 22266 | 0.00153 | 0.564 | Ã,Â¡ | 1 | = | 432 |

**Table 2**: Properties of different communication systems considered as the union of all messages expressed in English, Spanish, computer programing code, and MIDI music.

Graphs in Figure 4 allow to compare different communication systems at their fundamental scale. English and Spanish's profiles are very similar. The most frequent symbol is the space ' ', revealing that this particular character is for these languages, more than an actual symbol, part of the protocol used to indicate the start and the end of words. Both profiles exhibit two clearly differentiable ranges of behavior: a first rank range where the profile's slope increases its negative value, and a second rank range where the log-log profile's slope keeps nearly constant until no additional symbol exist and the frequency profile drops suddenly. Even though programming code and MIDI music exhibit a softer transition between these phases of behavior, they do show changes in their profile shapes according to the range of symbol ranking where it is observed.

For natural languages, English and Spanish, the transition between the two profile ranges appears as a nearly straight segment connecting them. Since single-character symbols fit in every place the character appears, they are useful to fill the interstice left in between longer symbols formed by several characters. Thus, it should be expected the communication system's alphabet and the punctuation signs to occupy, most of the head of the frequency distribution range, leaving the range of the tail for the longer and less frequent symbols. To ease the visualization of this effect, Figure 4 shows tags with the lowest ranked single-character symbols as well as the highest ranked multi-char symbols. Notice how these tags indicate the location of the profile's transition for English, Spanish and programming code, suggesting that the change of profile behavior is related to the number of characters forming each symbol.

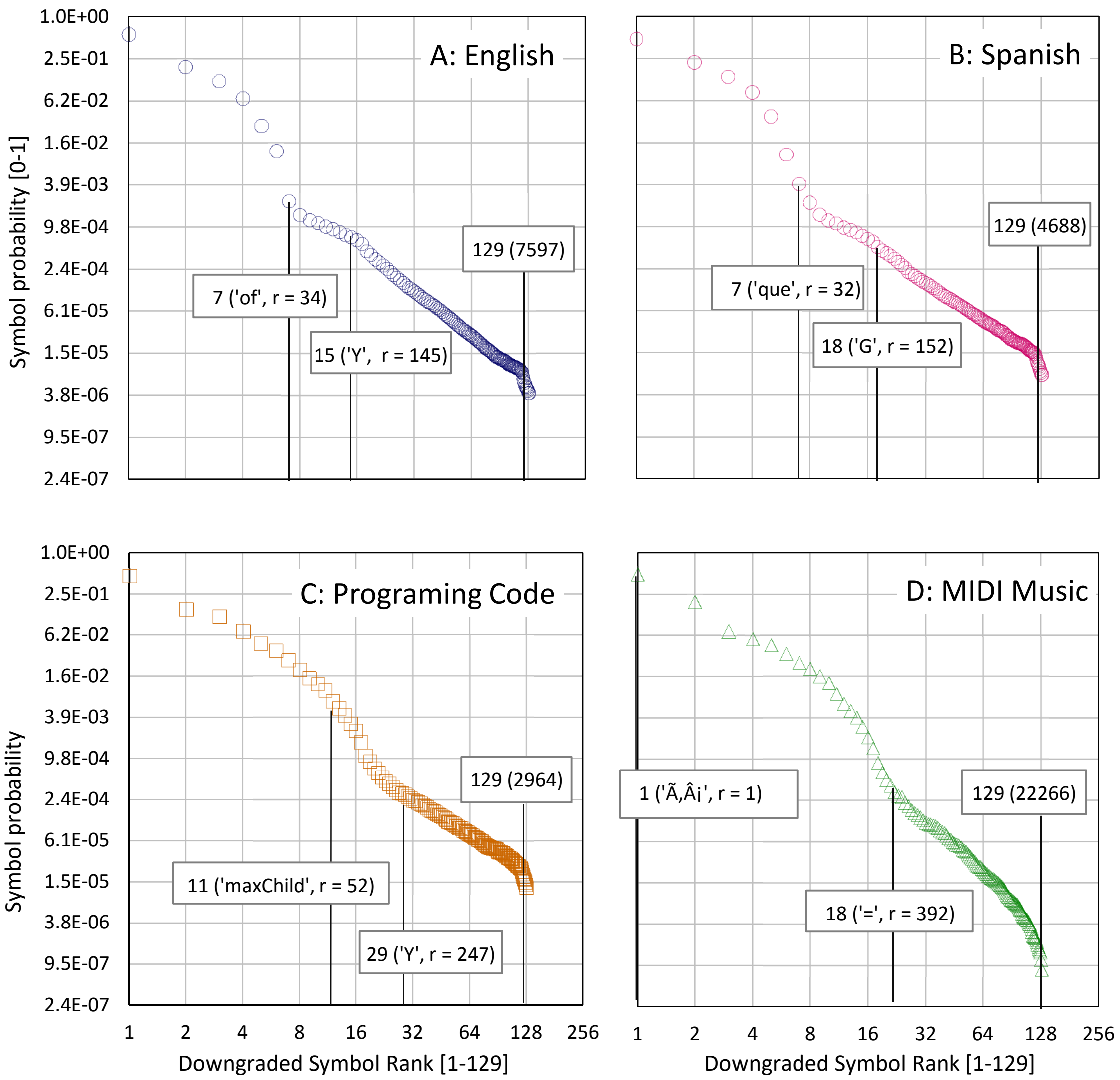

**Figure 4.** Probability profiles for several communication systems. A: English, B: Spanish, C: Programming code, D: MIDI music. All symbol rank axes are downgraded to the scale of 129. Numbers in tags show the symbol rank of the first multi-char. symbol and the last single-char. symbol as well as their corresponding symbol rank at the original scale (before downgrading). At the right end of each profile, the tags show the downgraded scale and the non-downgraded (original) scale.

The profile for MIDI music, represented in Figure 4D, shows a different behavior from those formerly viewed. For music the tendency of single-character symbols to occupy the heist ranked positions is not as dominant as it is for the natural languages and programming code.

In fact the most frequent symbol —the symbol ranked as r = 1— is the 4-chararacter symbol 'Ã‚Â¡'. Thus, the MIDI music profile starts right away with the transition from a phase with only single-character symbols, which does not manifest, to a phase dominated by longer and less frequent symbols at the profile's tail. For music as represented in computer files, there is no alphabet. Single characters symbols are not limited to the 26 or 28 letters of any alphabet. MIDI files, on the contrary, employ about 400 characters available in the Unicode character set. This explains why the transition range for music, ending with the least frequent single-char. Symbol, is around the 400th ranked symbol.

## 3.4 Stabilization Length

Figure 5 shows the values of entropy for the communication systems studied. English, Spanish and programming code are observed at the word, character and fundamental scales. MIDI music is observed at the character and the fundamental scale. Graphs included in Figure 5 show how entropy at all scales tend to decrease with the description length. For character and word scales, entropy seems to diminish indefinitely. At the fundamental scale all communication systems require some text length in order to '*develop*' the value of entropy. There appears to an asymptotic value at which the entropy to settles. We will call the entropy stabilization value $h_s$.

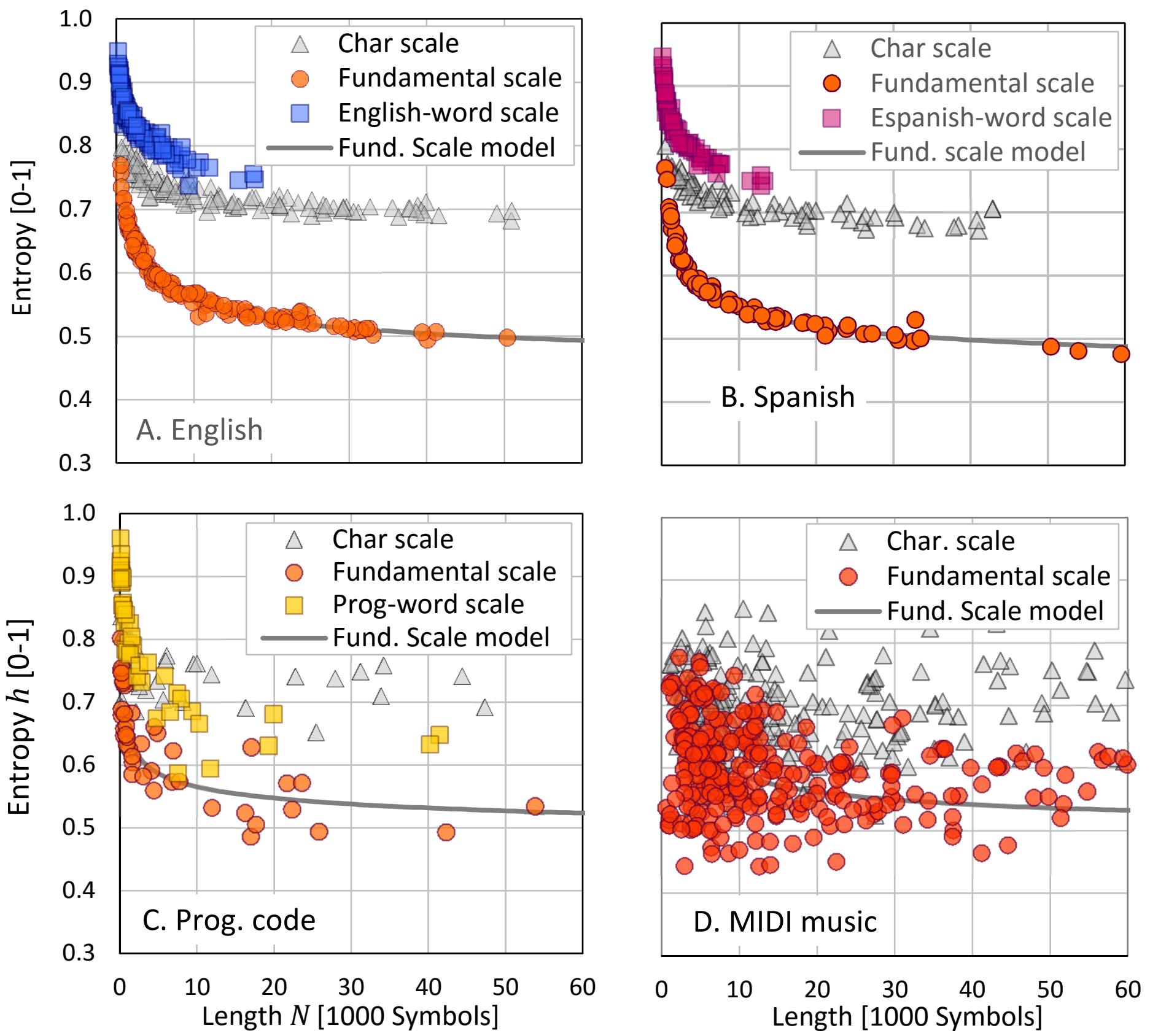

**Figure 5.** Entropy $h$ vs description length $N$ in characters. Graphs show the relationship between entropy and length for descriptions expressed in several types of communication systems: A. English, B: Spanish. C: Computer programing code and D: *MIDI* music.

In order to estimate the stabilization value, we built models of entropy as a function of the message length $N$ measured in symbols

$$h \approx 1 - h_s + \frac{1}{\mu \cdot N^{\nu}} \quad . \tag{7}$$

The parameters $\mu$ and $\nu$ are adjusted to minimize the error respect the real values presented in Figure 5. The values determined for the best minimal squared error fit at the fundamental scales are the following:

| | | | |
|---|---|---|---|
| English: | $h_s = 0.421$ | $\mu = 0.301$ | $\nu = 0.348$ |
| Spanish: | $h_s = 0.419$ | $\mu = 0.315$ | $\nu = 0.348$ |
| Programing code: | $h_s = 0.439$ | $\mu = 0.997$ | $\nu = 0.225$ |
| MIDI music: | $h_s = 0.479$ | $\mu = 0.213$ | $\nu = 0.407$ |

Figure 6 shows the expected entropy values from very short messages to the range of long messages, where expected entropy values become almost static. The rate at which the expected entropy approximates the established value $h_s$, is an indication of the length needed for a communication system to organize itself and reduce the entropy to convey the message. We arbitrarily set the lower limit of this range as the length at which the communication system's complexity reaches 80% of its settlement value. We refer to that value as the Stabilization Length $N_s$ and we measure it in characters. Once the considered stable range of entropy is numerically defined, the communication systems can be characterized by specific diversity and entropy found within their respective ranges.

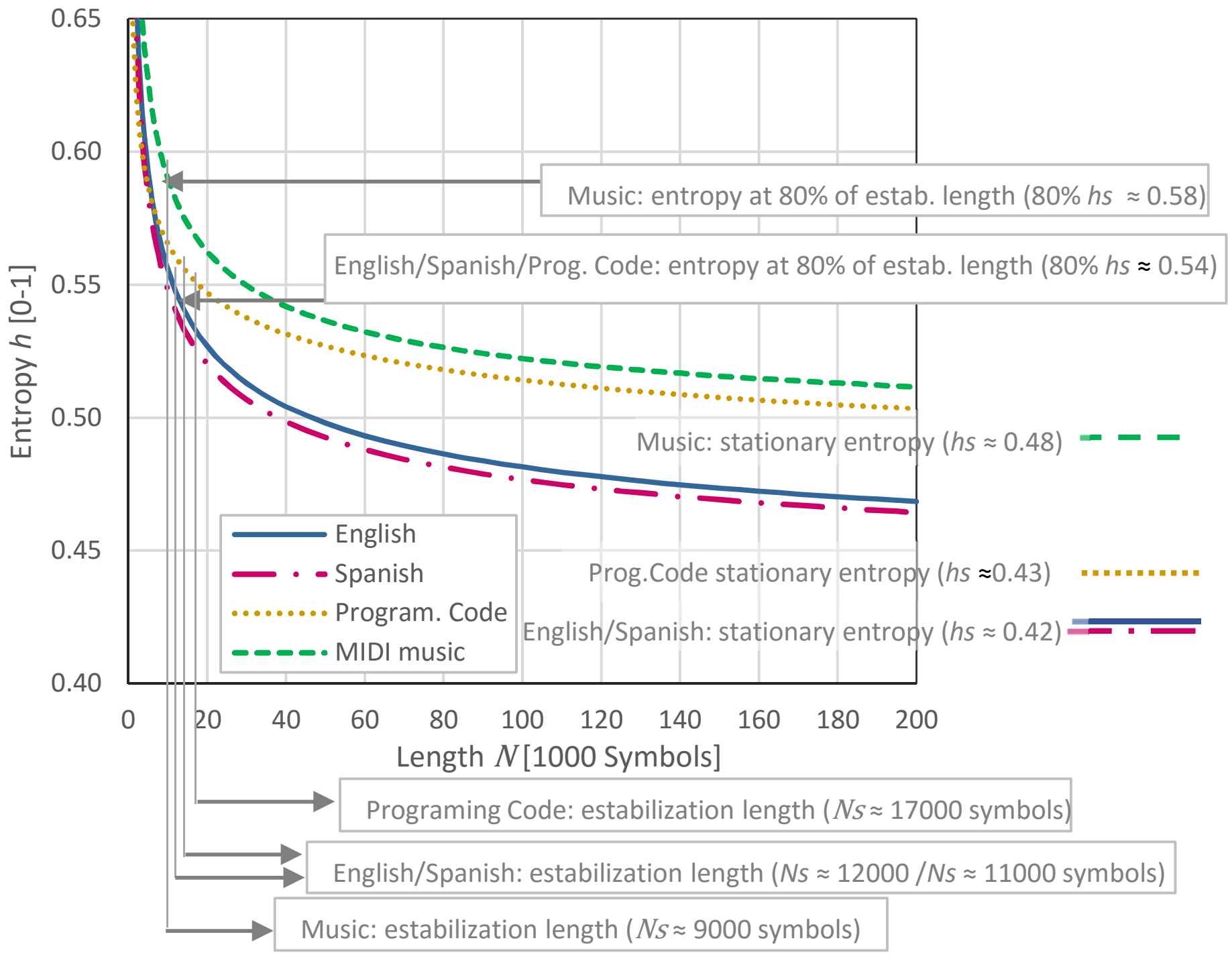

**Figure 6.** Model of entropy $h$ vs description length $N$ in characters. Graphs show the relationship between entropy and length for descriptions expressed in several types of communication systems.

The results of this characterization are included in Table 3. The Student t-tests p-values indicate that only for English and Spanish, their specific diversity and entropy of could come from similar distributions. All other combination of communication systems studied, are definitively different since the null hypothesis is discarded by the Student t-tests. The upper section of Table 3 includes averages and standard deviations for the specific diversity $d$ and the entropy $h$ of the communication systems studied. We shall highlight how smaller the standard deviation for the entropy of natural languages is as compared to the entropy's standard deviation of music and artificial languages.

Specific diversity and entropy for several types of communication systems at their fundamental scales

| Language Type | Specific Language | Stabilization length $N_s$ [symbols] | specific diversity $d$ Average | specific diversity $d$ Std. Dev. | entropy $h$ Average | entropy $h$ Std. Dev.* |
|---|---|---|---|---|---|---|
| Human natural | **English** | 12000 | 0.0460 | 0.00720 | 0.5239 | 0.0066 |
| | **Spanish** | 11000 | 0.0459 | 0.00755 | 0.5167 | 0.0060 |
| Artificial | **Comp. prog. Code** | 17000 | 0.0321 | 0.00824 | 0.5173 | 0.0221 |
| Music | ***MIDI* music** | 9000 | 0.0398 | 0.01266 | 0.5732 | 0.0517 |

Student t-tests

| Distributions 1-2 | $n_{s1}$ $n_{s2}$ | specific diversity $d$ p-value | entropy $h$ p-value |
|---|---|---|---|
| **English Spanish** | 47 32 | 0.9598 | 0.0946 |
| **English Comp.prog.** | 47 11 | **0.0011** | 0.4666 |
| **English Music** | 47 190 | **< 0.0001** | **< 0.0001** |
| **Spanish Comp.prog.** | 32 11 | **0.0011** | 0.9534 |
| **Spanish Music** | 32 190 | **0.0005** | **< 0.0001** |
| **Comp. prog. Music** | 11 190 | **0.0324** | **0.0001** |

* Standard deviation for entropy $h$ is calculated with respect to the value returned by the Fundamental Scale entropy model (Equation 7) at each description length $L$ measured in symbols.

**Table 3**. Average and standard deviation of the specific diversity, entropy and complexity for different types of communication systems, measured at the Fundamental Scale.

## 4. DISCUSSIONS

Human natural languages like English, Spanish or Chinese are syntactical. In their written form, they are constructed with symbols with an assigned meaning. The meaning of a symbol may vary slightly from an interpreter to another. And it may slowly change with time. In fact, the symbol itself can even become obsolete and disappear completely from the active version of the language.

Music language, in contrast, is made out of the effects obtained with combinations of sounds produced at different pitches, durations and time phases [10]. In spite of the audible essence of music, the sounds it is made of, can be coded as texts. In the context of this study we consider those music text-codes as musical language. Whether this conception of musical language actually represents the musical phenomena, is one the questions this study intends to answer.

Artificial languages, on the other hand, are represented in this study as algorithms coded in different programming languages. Computer programing languages are designed to give instructions to machines, but they are constructed using human natural languages symbols to produce the structures which ca be designed by humans. Computer programing languages are therefore a sort of instructional language based on natural language symbols. Artificial languages are specialized but very precise.

### 4.1. Diversity and Entropy

Figure 1 shows diversity as a function of message length. For natural languages at the fundamental scale this function exhibits very little deviations, suggesting that at fundamental scale the diversity is a function almost exclusively dependent of the description's length. Similarly, Figure 2 shows fundamental symbol entropy as a function specific diversity. Again, for natural languages the entropy is dominated by the specific diversity.

At the scale of words, Spanish shows a slightly lower entropy than English; confirming previous results presented in [6]. But at the fundamental scale and at character scale, English and Spanish do not show any important difference. For programing code the diversity, as well as the entropy, both measured at the scale of words, show high dispersion. An indication that words have little or no meaning for this kind of communication system. At its fundamental scale the dispersion of entropy reduces considerably but still is high compared to its counterpart for natural languages. The mix of many different programming languages in the same category may be an explanation. There is no word's scale for music. At its fundamental scale MIDI music shows the lowest specific diversity of all communication systems studied. This may be due to the nature of *MIDI* coding which in fact simplifies information during the digitizing process which basically consists of limiting the diversity of symbols associated to sound spectrum.

The values of diversity and the entropy computed at the fundamental scale, do not carry any distortion that may have been introduced by assuming a size of the scale, as is the case of the character's scale, or by assuming a symbol delimiter, as is the case of the scale of words. Graphs of diversity and entropy at the fundamental scale have been included In Figure 4 and Figure 5 to highlight the differences of those communication system's properties at that scale.

For English and Spanish at their fundamental scales, the symbol entropy is proportional to the specific diversity. For music, on the other hand, entropy shows more dispersion. Perhaps a consequence of the diversity of music types included in the study, which may behave like having several subtypes of communication systems into the same group.

The diversity of natural languages grows with the message length, behaving with small dispersion around an average value which follows the Heaps law; as expected. For computer programing code, the diversity is definitively lower than the diversity for natural languages. But *MIDI* music, as a communication system, while exhibits a large dispersion of symbolic diversity values, shows its cability to incorporate much richer variety of symbols than any other of the communication systems studied here. Entropy dispersion is conspicuously low in natural languages compared to artificial language and music. This hints to special structure or order in natural languages that is absent in the other two communication systems.

As signaled in Table 1, we included different genres and styles of music; from all over the world and covering more than 700 years of music transformation. Thus the *MIDI* music set studied is itself, a very diverse data set. As with the computer programing codes, here we could regard our set of musical pieces as expressions of many musical sublanguages, and therefore, considerable deviation should be expected for most communication system properties studied. However, there must be other important sources of deviations for the values of communication system properties, since for English and Spanish, in spite of being different expressions of natural languages, the fundamental scale showed overlapping curves of diversity and entropy, as if they were communication systems structurally equivalent.

### 4.3 Symbol Frequency Profiles

The number of characters of the symbols, combined with their relative frequency, are definitively related to the shape of the symbol frequency profile. While in the most frequent symbol region —the head of the ordered distribution— the increasingly negative slope corresponds to a Gaussian distribution of the symbols frequency, the tail of the distribution —where syllabus, word segments, complete words and other multi-character symbols appear— shows a power-law distribution resembling the qualitative profile shape announced by Zipf [8] in his early work.

### 4.2. Description Length.

Languages evolve to respond to '*stimuli*' of different kind exerted by the environment. For a human natural language for example, it is intuitively clear how people are prone to use more frequently those words which are written and pronounced using less space and time; this effect is the main argument behind the Zipf's principle of least effort [8] and the readability formulas for English [9] and Spanish [10].

The connection between word-entropy and readability for English and Spanish was explored by Febres and Jaffe [7]. Their findings signal a relationship between average word-length and sentence-length and symbolic entropy. Moreover, Febres and Jaffe show this numeric relationship goes beyond a mere quantitative effect and actually represents the possibility for evaluating styles of writing.

Of all communication systems studied here, English and Spanish show the lowest symbolic entropy. Reinforcing the idea that natural languages have evolved to be effective using resources in the transmission process, and being effective in the mutual understanding and coherence of the information shared by emitter and receiver. On the other side, MIDI music shows a more entropic distribution of symbols. This implies less compact messages. Music patterns are interconnected in such a way that anyone can detect a sound that does not correspond to a melody or polyphonic set of sounds. Even when listening a musical work for first time, a reasonably trained ear person may have a precise idea of what the short-time horizon sounds are expected. This effect of music language may explain why the complexity establishment length is so little for music; most music pieces expose their main theme rapidly; in very few compasses. Thus, the musical communication system requires shorter string lengths to develop its message.

For music the compactness of the message is not as important as it is for human natural languages. Music pursues other objectives not necessarily constrained in time and text length as English and Spanish do. Thereof, musical expressions have evolved not to be short but to produce certain feelings and sensations.

Entropy stabilization value $h_s$ is approximately 0.45 (range 0.419 to 0.479) suggesting an optimal complexity for all communication systems studied.

### 4.3. About the forces shaping languages

Our results show a connection among communication system properties as specific diversity and entropy. Observing the description at their fundamental scale, and obtaining the set of fundamental symbols, we were able to calculate the characteristic properties of communication systems presented in Table 2.

When focusing in the entropy of messages written in English and Spanish as represented in Figures 5 and 6, it can be seen that Spanish has a slightly lower entropy when compared with English. A result that suggests that Spanish, in spite of being formed by a number of words representing only a fraction of English words, is a more structured language. This is a consistent result with those obtained when language complexity was compared at the scale of words in [6], and with results shown in Figures 3 and 5 at the scale of words, where entropy values indicate a little more order for Spanish than for English. Explaining this slight but consistent difference between the complexities of English and Spanish, requires the inclusion of rigorous linguistics analysis and it lies beyond the purpose of this study. Nevertheless, we are tempted to mention that Spanish is the result of diversification process from Greek and Latin from which it inherited parts of its complex grammar from Latin. On the other hand, Modern English is the result of a conjunction of many dialects and old languages. Its evolution is then, characterize by its capacity to borrow words and to simplify, or to lose, grammar structures.

Natural languages and artificial languages have evolved to transfer and to record, precise complex ideas, the former, and precise instructions the latter. The effectiveness of both types of language rely on the presence of symbols with preconceived and shared meaning articulated by complex grammar rules to ensure the description in the contexts of place, time, actions, conditions, and all other elements that contribute to specify an idea. This functionality, together with the time they have had to evolve, explains the consistency of the entropy values obtained for natural languages.

In contract, musical language uses its capacity to trigger emotions and sensations rather than to convey concepts with preconceived meaning. Music is perceived as sequences of sounds patterns. Even though an almost unlimited sort of sounds can be incorporated to music, and explicit rules govern the essence of music and must be present in any pattern of sounds, for it to be considered as music. But in music the meaning of sounds or patterns of sounds do not have to be predefined. Certainly, there is a connection between sounds, harmonies and music scales with emotions. But that is not the result of a conscious and rational decision; any listener is free to feel and interpret music in a particular and personal fashion.

Having a different function from that of the natural languages, music is not anchored to keep its consistency as time passes; there is no meaning nor structure that music as a language has to maintain for long periods of time or large geographical areas. We think this *freedom* of music, specially manifested during the last two centuries, is the major factor that explains the vast variety of musical classes, genres, styles and even music definitions. Yet, within any branch of the music 'tree' at any time and region, music, as an audible phenomena, must obey a rather rigid network of relationships among its symbols which perhaps bounds the possibility from music being even more complex than it already is. In any case, Figure 5 shows how music exhibit a wide range of entropy values, at any range of the description length. Music initially results from the composer's feelings and inspiration; the composer *designs* his or her music to produce the desired emotions from the pattern of sounds. After being written in the music record, music tend to stick to the established sound structure defined as the style or genre. In music this structure seems to be governed by precise mathematical relations of sound duration and sound frequencies within the rhythms and superimposed accords which make polyphony music. When an instrument is played at an improper time or at an improper pitch, or plays an improper accord, the sound is considered to lose its beauty and in fact it may cause uncomfortable sensations for most listeners [11,12]. Yet some space remains free for the interpreter to alter the sound strictly described in the original music sheet, thereof any different interpretation of a musical piece adds —or subtracts— information to the musical description. Thus, the resulting entropy of a musical piece results from a personal way of using the language, initially imposed by the composer and then adjusted by the musicians who play the piece.

## 5. CONCLUSIONS

The character and word scales are the way we understand human natural languages; they allow us to learn and teach about them. The character scale and the word scale let us to organize complex languages into manageable components, but those scales do not seem to be the way languages, as adaptive entities, organize themselves. On the contrary, the symbols forming a fundamental scale, while being a difficult to set to determine, reveal much of the essence of each language and is a good basis to establish comparisons among languages of different types.

The fundamental scale is formed by those symbols having a dominant role within a description. Being the result of a computation with no assumptions about the size or delimiters of symbols, and being capable of representing the original description at a minimal length, the Fundamental Scale represents the best single-scale to study one-dimensional languages. Other observation scales may alter the evaluation of languages with the assumptions about their scale and structure, and thus results could be biased or misleading.

Human natural languages have evolved to transmit complex description in a precise and organized way. Being breve without diminishing content and precision, have been always an important aspect of the symbol generation and survival in the for natural language evolution process. This principles, captured in Zipf's law and Flesch's readability formulas, have molded natural languages up to their current status.

Natural languages are more symbolic diverse than music. But natural languages are dominated by grammar in a degree so high, that they show very thin dispersion around average property values. Music language, in its written form has a very limited number of symbols. Yet, due to the variations introduced when music is played, the assembly of sounds which constitute polyphonic music, the different instruments timbers, the rhythm syncopation, and many other effects of music as it sounds, the resulting symbolic diversity of music is, by a wide difference, the highest of all the languages studied.

The objective of music is not evolve to be effective in the sense of transmitting information. It probably evolves with another underlying sense of beauty, very difficult to describe in a quantitative manner. However, there is a fundamental scale for the music language which drives the sound patterns to constitute music. The possibility of knowing about the fundamental scale for specific music types, allows for a deeper studies of music as a language and detailed comparisons of the different types and styles with which music can be written and played.

Finally, being complexity dependent on entropy, an optimal complexity for all communication systems might exist.

# APPENDIX A

Example Text: symbol sets at different scales.

-What is an adverb? An adverb is a word or set of words that modifies verbs, adjectives, or other adverbs. An adverb answers how, when, where, or to what extent, how often or how much (e.g., daily, completely). Rule 1. Many adverbs end with the letters "ly", but many do not. An adverb is a word that changes or simplifies the meaning of a verb, adjective, other adverb, clause, or sentence expressing manner, place, time, or degree. Adverbs typically answer questions such as how?, in what why?, when?, where?, and to what extent?. Adverbs should never be confused with verbs. While verbs are used to describe actions, adverbs are used describe the way verbs are executed. Some adverbs can also modify adjectives as well as other adverbs.

$F_Y$ Frequency $E_Y$ Space occupied $N$ Message length [symbols] ø = space = max. symbol length = 13

| Chars Scale | | | Word Scale | | | | | | Fundamental Scale | | | | | | | |
|---|---|---|---|---|---|---|---|---|---|---|---|---|---|---|---|---|
| $D = 38$ | $h = 0.8080$ | | Diversity $D = 82$ | | | Entropy $h = 0.9033$ | | | Diversity $D = 80$ | | | | Entropy $h = 0.7628$ | | | |
| $d = 0.0$ | $N$ | 782 | Spec. diversity $d = 0.7033$ | | | Length $N$ | | 171 | Spec. diversity $d = 0.1384$ | | | | Length $N$ | | 578 | |
| Index | Symbol | $F_Y$ | Index | Symbol | $F_Y$ | Index | Symbol | $F_Y$ | Index | Symbol | $F_Y$ | $E_Y$ | Index | Symbol | $F_Y$ | $E_Y$ |
| 1 | ø | 169 | 1 | , | 21 | 41 | completely | 1 | 1 | ø | 100 | 1 | 41 | ul | 2 | 2 |
| 2 | e | 86 | 2 | . | 11 | 42 | ) | 1 | 2 | e | 70 | 1 | 42 | wi | 2 | 2 |
| 3 | a | 45 | 3 | or | 7 | 43 | Rule | 1 | 3 | a | 40 | 1 | 43 | io | 2 | 2 |
| 4 | s | 44 | 4 | adverbs | 7 | 44 | 1 | 1 | 4 | s | 36 | 1 | 44 | ie | 2 | 2 |
| 5 | r | 44 | 5 | ¿ | 6 | 45 | end | 1 | 5 | t | 36 | 1 | 45 | im | 2 | 2 |
| 6 | t | 39 | 6 | adverb | 5 | 46 | letters | 1 | 6 | r | 33 | 1 | 46 | whe | 2 | 3 |
| 7 | o | 34 | 7 | verbs | 4 | 47 | ly | 1 | 7 | o | 22 | 1 | 47 | øan | 2 | 3 |
| 8 | d | 32 | 8 | how | 4 | 48 | but | 1 | 8 | n | 21 | 1 | 48 | dif | 2 | 3 |
| 9 | n | 30 | 9 | an | 4 | 49 | do | 1 | 9 | , | 18 | 1 | 49 | uch | 2 | 3 |
| 10 | h | 28 | 10 | what | 4 | 50 | not | 1 | 10 | h | 17 | 1 | 50 | ,øc | 2 | 3 |
| 11 | i | 25 | 11 | is | 3 | 51 | changes | 1 | 11 | b | 12 | 1 | 51 | anyø | 2 | 4 |
| 12 | v | 21 | 12 | a | 3 | 52 | simplifies | 1 | 12 | dv | 10 | 2 | 52 | wordø | 2 | 5 |
| 13 | b | 21 | 13 | other | 3 | 53 | meaning | 1 | 13 | d | 9 | 1 | 53 | describ | 2 | 7 |
| 14 | w | 21 | 14 | to | 3 | 54 | verb | 1 | 14 | c | 8 | 1 | 54 | .øAdverb | 2 | 8 |
| 15 | , | 21 | 15 | the | 3 | 55 | adjective | 1 | 15 | u | 7 | 1 | 55 | ød | 1 | 2 |
| 16 | c | 17 | 16 | as | 3 | 56 | clause | 1 | 16 | l | 6 | 1 | 56 | øv | 1 | 2 |
| 17 | l | 16 | 17 | are | 3 | 57 | sentence | 1 | 17 | ¿ | 6 | 1 | 57 | word | 1 | 4 |
| 18 | . | 11 | 18 | word | 2 | 58 | expressing | 1 | 18 | wh | 6 | 2 | 58 | yø | 1 | 2 |
| 19 | u | 11 | 19 | of | 2 | 59 | manner | 1 | 19 | w | 5 | 1 | 59 | ma | 1 | 2 |
| 20 | m | 10 | 20 | that | 2 | 60 | place | 1 | 20 | i | 5 | 1 | 60 | f | 1 | 1 |
| 21 | y | 10 | 21 | adjectives | 2 | 61 | time | 1 | 21 | . | 4 | 2 | 61 | ns | 1 | 2 |
| 22 | f | 7 | 22 | when | 2 | 62 | degree | 1 | 22 | g | 4 | 1 | 62 | An | 1 | 2 |
| 23 | ¿ | 6 | 23 | where | 2 | 63 | typically | 1 | 23 | x | 4 | 1 | 63 | w | 1 | 2 |
| 24 | A | 5 | 24 | extent | 2 | 64 | answer | 1 | 24 | ly | 4 | 2 | 64 | b, | 1 | 2 |
| 25 | g | 5 | 25 | with | 2 | 65 | questions | 1 | 25 | m | 4 | 1 | 65 | v | 1 | 1 |
| 26 | p | 5 | 26 | " | 2 | 66 | such | 1 | 26 | verbs | 4 | 5 | 66 | - | 1 | 1 |
| 27 | x | 4 | 27 | used | 2 | 67 | in | 1 | 27 | y | 3 | 1 | 67 | ( | 1 | 1 |
| 28 | j | 3 | 28 | describe | 2 | 68 | why | 1 | 28 | p | 3 | 1 | 68 | ) | 1 | 1 |
| 29 | W | 2 | 29 | many | 2 | 69 | and | 1 | 29 | dj | 3 | 2 | 69 | R | 1 | 1 |
| 30 | " | 2 | 30 | - | 1 | 70 | should | 1 | 30 | øof | 3 | 3 | 70 | I | 1 | 1 |
| 31 | - | 1 | 31 | set | 1 | 71 | never | 1 | 31 | ctiv | 3 | 4 | 71 | M | 1 | 1 |
| 32 | ( | 1 | 32 | words | 1 | 72 | be | 1 | 32 | .øA | 2 | 3 | 72 | q | 1 | 1 |
| 33 | ) | 1 | 33 | modifies | 1 | 73 | confused | 1 | 33 | . | 2 | 1 | 73 | S | 1 | 1 |
| 34 | R | 1 | 34 | answers | 1 | 74 | While | 1 | 34 | W | 2 | 1 | 74 | ho | 1 | 2 |
| 35 | I | 1 | 35 | often | 1 | 75 | actions | 1 | 35 | " | 2 | 1 | 75 | øm | 1 | 2 |
| 36 | M | 1 | 36 | much | 1 | 76 | way | 1 | 36 | ow | 2 | 2 | 76 | ng | 1 | 2 |
| 37 | q | 1 | 37 | € | 1 | 77 | executed | 1 | 37 | me | 2 | 2 | 77 | it | 1 | 2 |
| 38 | S | 1 | 38 | e | 1 | 78 | Some | 1 | 38 | le | 2 | 2 | 78 | in | 1 | 2 |
| | | | 39 | g | 1 | 79 | can | 1 | 39 | øi | 2 | 2 | 79 | on | 1 | 2 |
| | | | 40 | daily | 1 | 80 | also | 1 | 40 | pl | 2 | 2 | 80 | si | 1 | 2 |
| | | | | | | 81 | modify | 1 | | | | | | | | |
| | | | | | | 82 | well | 1 | | | | | | | | |

# APPENDIX B

EnglishProperties.htm

## English Properties at Different Scales

*L = Length* — *d = Specific Diversity* — *h = entropy*

*c = Complexity* — *[F.S.] = [Fundamental Symbols]* — *[chrs] = [characters]*

*[w] = [words]* — *[0-1] = between 0 and 1*

| | At Char Scale | | | At Fundamental Scale | | | At Word Scale | | |
|---|---|---|---|---|---|---|---|---|---|
| | *L* | *d* | *h* | *L* | *d* | *h* | *L* | *d* | *h* |
| *Message Name* | [chrs] | [0-1] | [0-1] | [F.S.] | [0-1] | [0-1] | [w] | [0-1] | [0-1] |
| 1381.JohnBall.txt | 1122 | 0.0294 | 0.830 | 899 | 0.121 | 0.715 | 227 | 0.515 | 0.914 |
| 1601.Hamlet.txt | 645 | 0.0220 | 0.792 | 563 | 0.124 | 0.758 | 150 | 0.435 | 0.950 |
| 1588.QueenElizabethI.txt | 1635 | 0.0636 | 0.815 | 1256 | 0.111 | 0.688 | 359 | 0.647 | 0.879 |
| 1601.QueenElizabethI.txt | 5451 | 0.0097 | 0.744 | 4275 | 0.078 | 0.607 | 1140 | 0.340 | 0.865 |
| 1851.SojournerTruth.txt | 1881 | 0.0271 | 0.748 | 1529 | 0.097 | 0.666 | 443 | 0.413 | 0.911 |
| 1877.ChiefJoseph.txt | 770 | 0.0532 | 0.786 | 603 | 0.126 | 0.736 | 183 | 0.503 | 0.926 |
| 1901.MarkTwain.txt | 2971 | 0.0162 | 0.755 | 2366 | 0.093 | 0.633 | 669 | 0.386 | 0.889 |
| 1923.BS.Eng.WilliamButlerYeats.txt | 1689 | 0.0243 | 0.791 | 1363 | 0.096 | 0.691 | 320 | 0.522 | 0.920 |
| 1932.MargaretSanger.txt | 6123 | 0.0085 | 0.737 | 4722 | 0.077 | 0.586 | 1162 | 0.343 | 0.847 |
| 1936.KingEdwardVIII.txt | 2850 | 0.0147 | 0.788 | 2208 | 0.098 | 0.653 | 596 | 0.408 | 0.875 |
| 1938.BS.PearlBuck.txt | 2458 | 0.0175 | 0.779 | 1911 | 0.094 | 0.660 | 520 | 0.379 | 0.893 |
| 1940.05.WinstonChurchill.txt | 3530 | 0.0150 | 0.744 | 2910 | 0.077 | 0.646 | 703 | 0.415 | 0.873 |
| 1941.FranklinDRoosvelt.txt | 3184 | 0.0173 | 0.747 | 2496 | 0.091 | 0.634 | 574 | 0.455 | 0.881 |
| 1942.MahatmaGandhi.txt | 6106 | 0.0097 | 0.725 | 4724 | 0.073 | 0.604 | 1234 | 0.347 | 0.855 |
| 1944.DwightEisenhower.txt | 1076 | 0.0418 | 0.788 | 885 | 0.122 | 0.718 | 208 | 0.577 | 0.925 |
| 1944.GeorgePatton.txt | 3919 | 0.0138 | 0.744 | 3026 | 0.074 | 0.639 | 890 | 0.361 | 0.886 |
| 1946.WinstonChurchill.txt | 6633 | 0.0089 | 0.725 | 5139 | 0.080 | 0.587 | 1285 | 0.388 | 0.850 |
| 1949.BS.Eng.WilliamFaulkner.t | 3016 | 0.0136 | 0.780 | 2414 | 0.088 | 0.646 | 622 | 0.399 | 0.884 |
| 1954.BS.Eng.ErnestHemingway | 1808 | 0.0227 | 0.777 | 1452 | 0.101 | 0.673 | 367 | 0.499 | 0.919 |
| 1961.11.JohnFKennedy.txt | 3596 | 0.0131 | 0.760 | 2816 | 0.091 | 0.629 | 680 | 0.465 | 0.892 |
| 1962.BS.Eng.JohnSteinbeck.txt | 4925 | 0.0091 | 0.769 | 3852 | 0.084 | 0.616 | 952 | 0.404 | 0.859 |
| 1963.06.26.JohnFKennedy.txt | 3146 | 0.0146 | 0.761 | 2452 | 0.082 | 0.633 | 665 | 0.358 | 0.875 |
| 1964.05.LyndonBJohnson.txt | 5975 | 0.0095 | 0.729 | 4742 | 0.072 | 0.600 | 1168 | 0.368 | 0.848 |
| 1964.LadybirdJohnson.txt | 4205 | 0.0150 | 0.719 | 3347 | 0.084 | 0.621 | 818 | 0.432 | 0.876 |
| 1964.MartinLutherKing.txt | 6667 | 0.0081 | 0.738 | 5006 | 0.076 | 0.592 | 1266 | 0.393 | 0.862 |
| 1968.RobertFKennedy.txt | 3025 | 0.0152 | 0.765 | 2353 | 0.080 | 0.636 | 629 | 0.315 | 0.898 |
| 1969.IndiraGhandi.txt | 5279 | 0.0100 | 0.746 | 3962 | 0.078 | 0.608 | 1058 | 0.386 | 0.867 |
| 1969.ShirleyChisholm.txt | 5102 | 0.0114 | 0.730 | 3983 | 0.078 | 0.602 | 967 | 0.396 | 0.867 |
| 1972.JaneFonda.txt | 4053 | 0.0136 | 0.744 | 3095 | 0.091 | 0.620 | 799 | 0.432 | 0.873 |
| 1976.BS.Eng.SaulBellow.txt | 1999 | 0.0255 | 0.758 | 1607 | 0.105 | 0.672 | 397 | 0.501 | 0.912 |
| 1981.RonaldReagan.txt | 5877 | 0.0095 | 0.732 | 4552 | 0.074 | 0.599 | 1183 | 0.383 | 0.856 |
| 1983.BS.Eng.WilliamGolding.tx | 1837 | 0.0256 | 0.772 | 1538 | 0.097 | 0.684 | 369 | 0.545 | 0.913 |
| 1986.BS.Eng.WoleSoyinka.txt | 2529 | 0.0178 | 0.781 | 1972 | 0.099 | 0.666 | 482 | 0.508 | 0.892 |
| 1986.RonaldReagan.txt | 3738 | 0.0158 | 0.732 | 3093 | 0.075 | 0.635 | 805 | 0.385 | 0.857 |
| 1991.BS.Eng.NadineGordimer.t | 2837 | 0.0176 | 0.756 | 2265 | 0.093 | 0.651 | 564 | 0.500 | 0.892 |
| 1992.BS.Eng.DerekWalcott.txt | 611 | 0.0704 | 0.798 | 545 | 0.127 | 0.770 | 104 | 0.654 | 0.930 |
| 1993.BS.Eng.ToniMorrison.txt | 1887 | 0.0228 | 0.783 | 1514 | 0.103 | 0.675 | 368 | 0.546 | 0.912 |
| 1993.MayaAngelou.txt | 3660 | 0.0145 | 0.757 | 3000 | 0.082 | 0.644 | 794 | 0.392 | 0.835 |

| | At Char Scale | | | At Fundamental Scale | | | At Word Scale | | |
|---|---|---|---|---|---|---|---|---|---|
| | *L* | *d* | *h* | *L* | *d* | *h* | *L* | *d* | *h* |
| *Message Name* | [chrs] | [0-1] | [0-1] | [F.S.] | [0-1] | [0-1] | [w] | [0-1] | [0-1] |
| 1993.SarahBrady.txt | 4409 | 0.0118 | 0.752 | 3555 | 0.069 | 0.618 | 969 | 0.332 | 0.869 |
| 1993.UrvashiVaid.txt | 6545 | 0.0089 | 0.737 | 4895 | 0.076 | 0.595 | 1319 | 0.315 | 0.841 |
| 1994.NelsonMandela.txt | 5181 | 0.0097 | 0.748 | 3930 | 0.043 | 0.633 | 1010 | 0.384 | 0.848 |
| 1995.BS.Eng.SeamusHeaney.tx | 1508 | 0.0312 | 0.769 | 1225 | 0.112 | 0.689 | 287 | 0.561 | 0.915 |
| 1997.BillClinton.txt | 6083 | 0.0087 | 0.741 | 4610 | 0.084 | 0.592 | 1303 | 0.322 | 0.845 |
| 1997.QueenElizabethII.txt | 2172 | 0.0203 | 0.771 | 1795 | 0.088 | 0.665 | 449 | 0.461 | 0.900 |
| 2001.09.11.GeorgeWBush.txt | 3522 | 0.0139 | 0.760 | 2777 | 0.086 | 0.637 | 673 | 0.443 | 0.882 |
| 2001.09.13.GeorgeWBush.txt | 2922 | 0.0198 | 0.740 | 2484 | 0.078 | 0.654 | 550 | 0.456 | 0.874 |
| 2001.BS.Eng.VSNaipaul.txt | 1665 | 0.0306 | 0.760 | 1365 | 0.113 | 0.679 | 348 | 0.500 | 0.899 |
| 2001.HalleBerry.txt | 2840 | 0.0204 | 0.748 | 2303 | 0.089 | 0.654 | 649 | 0.337 | 0.849 |
| 2002.OprahWinfrey.txt | 2685 | 0.0175 | 0.769 | 2061 | 0.099 | 0.644 | 609 | 0.373 | 0.865 |
| 2003.BethChapman.txt | 4257 | 0.0148 | 0.720 | 3452 | 0.074 | 0.620 | 882 | 0.382 | 0.877 |
| 2003.BS.Eng.JMCoetzee.txt | 1510 | 0.0364 | 0.757 | 1239 | 0.098 | 0.700 | 331 | 0.459 | 0.913 |
| 1606.LancelotAndrewes.txt | 41451 | 0.0017 | 0.691 | 32985 | 0.040 | 0.503 | 9291 | 0.166 | 0.738 |
| 1833.ThomasBabington.txt | 81977 | 0.0009 | 0.688 | 62980 | 0.035 | 0.487 | 15668 | 0.169 | 0.746 |
| 1849.LucretiaMott.txt | 38756 | 0.0017 | 0.707 | 30664 | 0.043 | 0.509 | 7577 | 0.227 | 0.770 |
| 1851.ErnestineLRose.txt | 39851 | 0.0016 | 0.711 | 32514 | 0.036 | 0.514 | 8301 | 0.196 | 0.764 |
| 1861.AbrahamLincoln.txt | 20952 | 0.0027 | 0.722 | 16550 | 0.051 | 0.537 | 4007 | 0.254 | 0.808 |
| 1867.ElizabethCadyStanton.txt | 29592 | 0.0022 | 0.705 | 23717 | 0.036 | 0.541 | 5862 | 0.253 | 0.784 |
| 1890.RusselConwell.txt | 81989 | 0.0009 | 0.686 | 63660 | 0.034 | 0.483 | 17795 | 0.128 | 0.748 |
| 1892.FrancesEWHarper.txt | 21988 | 0.0026 | 0.719 | 17224 | 0.051 | 0.537 | 4396 | 0.283 | 0.805 |
| 1906.MaryChurch.txt | 8158 | 0.0072 | 0.722 | 6570 | 0.070 | 0.577 | 1558 | 0.375 | 0.852 |
| 1909.BS.SelmaLagerlof.txt | 10046 | 0.0061 | 0.715 | 8424 | 0.059 | 0.568 | 2301 | 0.272 | 0.826 |
| 1915.AnnaHoward.txt | 50806 | 0.0014 | 0.683 | 40013 | 0.036 | 0.496 | 10652 | 0.134 | 0.776 |
| 1916.CarrieChapman.txt | 31123 | 0.0023 | 0.696 | 24697 | 0.047 | 0.521 | 6127 | 0.252 | 0.794 |
| 1916.HellenKeller.txt | 13143 | 0.0046 | 0.724 | 10498 | 0.081 | 0.532 | 2562 | 0.335 | 0.829 |
| 1918.WoodrowWilson.txt | 15039 | 0.0043 | 0.702 | 12279 | 0.050 | 0.555 | 2753 | 0.279 | 0.818 |
| 1920.CrystalEastman.txt | 10557 | 0.0051 | 0.733 | 8326 | 0.071 | 0.564 | 2136 | 0.314 | 0.848 |
| 1923.JamesMonroe.txt | 6485 | 0.0076 | 0.743 | 5151 | 0.067 | 0.596 | 1178 | 0.354 | 0.849 |
| 1923.NL.Eng.WilliamButlerYea | 21120 | 0.0031 | 0.704 | 16724 | 0.053 | 0.539 | 4258 | 0.265 | 0.819 |
| 1925.MaryReynolds.txt | 17911 | 0.0031 | 0.719 | 14404 | 0.059 | 0.535 | 4340 | 0.198 | 0.799 |
| 1930.NL.Eng.SinclairLewis.txt | 29220 | 0.0023 | 0.705 | 24545 | 0.040 | 0.535 | 5708 | 0.282 | 0.799 |
| 1936.EleanorRoosvelt.txt | 9186 | 0.0063 | 0.710 | 7082 | 0.062 | 0.573 | 1968 | 0.233 | 0.830 |
| 1938.NL.PearlBuck.txt | 50855 | 0.0013 | 0.698 | 41104 | 0.033 | 0.507 | 10270 | 0.178 | 0.767 |
| 1940.06.A.WinstonChurchill.tx | 19584 | 0.0035 | 0.707 | 15511 | 0.052 | 0.545 | 3784 | 0.282 | 0.822 |
| 1940.06.B.WinstonChurchill.tx | 25152 | 0.0025 | 0.715 | 20006 | 0.049 | 0.529 | 4909 | 0.242 | 0.802 |
| 1941.HaroldIckes.txt | 12131 | 0.0046 | 0.736 | 9806 | 0.060 | 0.568 | 2449 | 0.295 | 0.822 |
| 1947.GeorgeCMarshall.txt | 8669 | 0.0058 | 0.746 | 6843 | 0.071 | 0.576 | 1608 | 0.363 | 0.843 |
| 1947.HarryTruman.txt | 13420 | 0.0045 | 0.720 | 11008 | 0.056 | 0.557 | 2459 | 0.292 | 0.821 |
| 1948.BS.Eng.ThomasEliot.txt | 7381 | 0.0078 | 0.724 | 5864 | 0.082 | 0.582 | 1467 | 0.344 | 0.845 |
| 1950.MargaretChase.txt | 9313 | 0.0056 | 0.749 | 7284 | 0.071 | 0.578 | 1717 | 0.327 | 0.845 |
| 1950.NL.Eng.BertrandRussell.t | 32621 | 0.0021 | 0.705 | 25362 | 0.049 | 0.522 | 5716 | 0.327 | 0.821 |
| 1953.DwightEisenhower.txt | 14887 | 0.0042 | 0.709 | 11441 | 0.059 | 0.549 | 2910 | 0.285 | 0.811 |
| 1953.NelsonMandela.txt | 27937 | 0.0025 | 0.703 | 22128 | 0.046 | 0.530 | 4967 | 0.289 | 0.801 |
| 1957.MartinLutherKing.txt | 39237 | 0.0018 | 0.700 | 31528 | 0.039 | 0.510 | 7953 | 0.159 | 0.780 |
| 1959.RichardFeynman.txt | 39621 | 0.0018 | 0.693 | 31335 | 0.042 | 0.511 | 8218 | 0.159 | 0.786 |

| | At Char Scale | | | At Fundamental Scale | | | At Word Scale | | |
|---|---|---|---|---|---|---|---|---|---|
| | *L* | *d* | *h* | *L* | *d* | *h* | *L* | *d* | *h* |
| *Message Name* | [chrs] | [0-1] | [0-1] | [F.S.] | [0-1] | [0-1] | [w] | [0-1] | [0-1] |
| 1961.01.JohnFKennedy.txt | 7433 | 0.0070 | 0.739 | 6039 | 0.070 | 0.588 | 1521 | 0.348 | 0.852 |
| 1961.04.JohnFKennedy.txt | 8697 | 0.0071 | 0.708 | 6936 | 0.068 | 0.568 | 1715 | 0.353 | 0.845 |
| 1961.05.JohnFKennedy.txt | 35545 | 0.0019 | 0.703 | 27984 | 0.048 | 0.517 | 6588 | 0.233 | 0.799 |
| 1962.09.JohnFKennedy.txt | 11652 | 0.0057 | 0.697 | 9748 | 0.061 | 0.554 | 2441 | 0.308 | 0.827 |
| 1962.10.JohnFKennedy.txt | 14787 | 0.0045 | 0.708 | 11833 | 0.056 | 0.552 | 2772 | 0.293 | 0.829 |
| 1962.12.MalcomX.txt | 80830 | 0.0009 | 0.697 | 62446 | 0.034 | 0.479 | 17561 | 0.095 | 0.757 |
| 1963.06.10.JohnFKennedy.txt | 18539 | 0.0035 | 0.699 | 14797 | 0.056 | 0.545 | 3680 | 0.277 | 0.815 |
| 1963.09.20.JohnFKennedy.txt | 20998 | 0.0031 | 0.706 | 16571 | 0.055 | 0.534 | 3988 | 0.273 | 0.804 |
| 1963.MartinLutherKing.txt | 8526 | 0.0063 | 0.736 | 6873 | 0.067 | 0.586 | 1731 | 0.304 | 0.837 |
| 1964.04.MalcomX.txt | 15616 | 0.0044 | 0.706 | 11543 | 0.074 | 0.537 | 3381 | 0.198 | 0.817 |
| 1964.NelsonMandela.txt | 63224 | 0.0012 | 0.699 | 50334 | 0.034 | 0.498 | 11935 | 0.180 | 0.767 |
| 1965.03.LyndonBJohnson.txt | 20217 | 0.0031 | 0.716 | 16523 | 0.049 | 0.544 | 4169 | 0.235 | 0.805 |
| 1965.04.LyndonBJohnson.txt | 6171 | 0.0088 | 0.733 | 4919 | 0.069 | 0.598 | 1286 | 0.326 | 0.849 |
| 1967.MartinLutherKing.txt | 37609 | 0.0018 | 0.702 | 30603 | 0.043 | 0.514 | 7366 | 0.237 | 0.794 |
| 1968.MartinLutherKing.txt | 23357 | 0.0028 | 0.711 | 18085 | 0.051 | 0.535 | 5119 | 0.195 | 0.793 |
| 1969.RichardNixon.txt | 26380 | 0.0027 | 0.700 | 20360 | 0.050 | 0.524 | 5070 | 0.218 | 0.805 |
| 1972.RichardNixon.txt | 25132 | 0.0030 | 0.690 | 19868 | 0.042 | 0.527 | 5406 | 0.172 | 0.795 |
| 1974.RichardNixon.txt | 9735 | 0.0056 | 0.730 | 7773 | 0.064 | 0.570 | 1959 | 0.274 | 0.833 |
| 1976.NL.Eng.SaulBellow.txt | 28967 | 0.0025 | 0.699 | 23205 | 0.046 | 0.527 | 5639 | 0.266 | 0.799 |
| 1979.MargaretThatcher.txt | 17079 | 0.0037 | 0.715 | 13812 | 0.057 | 0.541 | 3219 | 0.312 | 0.820 |
| 1982.RonaldReagan.txt | 26695 | 0.0027 | 0.695 | 21482 | 0.046 | 0.531 | 5052 | 0.277 | 0.807 |
| 1983.NL.Eng.WilliamGolding.tx | 24742 | 0.0026 | 0.703 | 20114 | 0.052 | 0.534 | 5150 | 0.267 | 0.812 |
| 1983.RonaldReagan.txt | 26631 | 0.0025 | 0.708 | 21269 | 0.052 | 0.528 | 5236 | 0.242 | 0.814 |
| 1986.NL.Eng.WoleSoyinka.txt | 48990 | 0.0015 | 0.695 | 39377 | 0.038 | 0.507 | 9034 | 0.280 | 0.778 |
| 1987.RonaldReagan.txt | 15753 | 0.0043 | 0.705 | 12717 | 0.057 | 0.553 | 3170 | 0.296 | 0.822 |
| 1988.AnnRichards.txt | 15083 | 0.0043 | 0.712 | 11992 | 0.057 | 0.558 | 3126 | 0.278 | 0.827 |
| 1991.GeorgeBush.txt | 8716 | 0.0064 | 0.739 | 7152 | 0.069 | 0.584 | 1782 | 0.328 | 0.844 |
| 1991.NL.Eng.NadineGordimer.t | 22521 | 0.0032 | 0.694 | 18005 | 0.052 | 0.533 | 4386 | 0.286 | 0.802 |
| 1992.NL.Eng.DerekWalcott.txt | 37759 | 0.0018 | 0.702 | 29694 | 0.046 | 0.512 | 7407 | 0.266 | 0.774 |
| 1993.NL.Eng.ToniMorrison.txt | 17471 | 0.0035 | 0.719 | 13811 | 0.059 | 0.549 | 3492 | 0.294 | 0.813 |
| 1995.ErikaJong.txt | 12131 | 0.0051 | 0.717 | 10030 | 0.049 | 0.569 | 2401 | 0.252 | 0.832 |
| 1995.HillaryClinton.txt | 12878 | 0.0048 | 0.714 | 10546 | 0.053 | 0.567 | 2487 | 0.289 | 0.823 |
| 1995.NL.Eng.SeamusHeaney.tx | 36355 | 0.0020 | 0.689 | 28865 | 0.045 | 0.517 | 7054 | 0.270 | 0.787 |
| 1997.EarlOfSpencer.txt | 6509 | 0.0078 | 0.741 | 5284 | 0.073 | 0.598 | 1327 | 0.383 | 0.857 |
| 1997.NancyBirdsall.txt | 13010 | 0.0052 | 0.717 | 10323 | 0.055 | 0.569 | 2312 | 0.279 | 0.833 |
| 1997.PrincessDiana.txt | 8558 | 0.0072 | 0.717 | 6942 | 0.071 | 0.576 | 1759 | 0.343 | 0.849 |
| 1999.AnitaRoddick.txt | 9966 | 0.0059 | 0.722 | 7987 | 0.072 | 0.568 | 2040 | 0.313 | 0.843 |
| 2000.CondoleezzaRice.txt | 7551 | 0.0079 | 0.729 | 5972 | 0.073 | 0.591 | 1511 | 0.341 | 0.854 |
| 2000.CourtneyLove.txt | 38575 | 0.0019 | 0.697 | 32153 | 0.041 | 0.513 | 8344 | 0.196 | 0.799 |
| 2001.NL.Eng.VSNaipaul.txt | 30520 | 0.0024 | 0.697 | 23708 | 0.047 | 0.524 | 6327 | 0.194 | 0.788 |
| 2003.NL.Eng.JMCoetzee.txt | 20992 | 0.0028 | 0.708 | 16858 | 0.052 | 0.531 | 4593 | 0.241 | 0.793 |
| 2005.NL.Eng.HaroldPinter.txt | 29213 | 0.0025 | 0.702 | 23586 | 0.041 | 0.539 | 5833 | 0.255 | 0.803 |
| 2005.SteveJobs.txt | 12200 | 0.0056 | 0.708 | 9360 | 0.065 | 0.568 | 2615 | 0.273 | 0.832 |
| 2007.NL.Eng.DorisLessing.txt | 26956 | 0.0024 | 0.706 | 21783 | 0.046 | 0.524 | 5898 | 0.212 | 0.793 |

SpanishProperties.htm

# Spanish Properties at Different Scales

*L = Length* *c = Complexity* *[w] = [words]*
*d = Specific Diversity* *[F.S.] = [Fundamental Symbols]*
*h = entropy* *[chrs] = [characters]* *[0-1] = between 0 and 1*

| | At Char Scale | | | At Fundamental Scale | | | At Word Scale | | |
|---|---|---|---|---|---|---|---|---|---|
| | *L* | *d* | *h* | *L* | *d* | *h* | *L* | *d* | *h* |
| *Message Name* | [chrs] | [0-1] | [0-1] | [F.S.] | [0-1] | [0-1] | [w] | [0-1] | [0-1] |
| 1805.Simon Bolivar.txt | 2480 | 0.0226 | 0.734 | 2118 | 0.0817 | 0.660 | 462 | 0.4978 | 0.878 |
| 1813.Simon Bolivar.txt | 4127 | 0.0119 | 0.749 | 3306 | 0.0811 | 0.612 | 739 | 0.4493 | 0.864 |
| 1830.Simon Bolivar.txt | 1120 | 0.0402 | 0.768 | 923 | 0.1073 | 0.709 | 201 | 0.6020 | 0.930 |
| 1931.Manuel Azana.txt | 1617 | 0.0291 | 0.763 | 1282 | 0.0983 | 0.675 | 297 | 0.5118 | 0.906 |
| 1936.Dolores Ibarruri.txt | 3069 | 0.0199 | 0.733 | 2204 | 0.0867 | 0.626 | 641 | 0.3276 | 0.863 |
| 1936.Jose Buenaventura Durru | 3672 | 0.0155 | 0.726 | 2977 | 0.0820 | 0.605 | 690 | 0.4420 | 0.877 |
| 1938.Dolores Ibarruri.txt | 4273 | 0.0133 | 0.727 | 3468 | 0.0819 | 0.597 | 774 | 0.4109 | 0.846 |
| 1945.Juan Domingo Perón.txt | 5861 | 0.0096 | 0.726 | 4760 | 0.0666 | 0.596 | 1192 | 0.3649 | 0.866 |
| 1959.Fulgencio Batista.txt | 466 | 0.0751 | 0.806 | 416 | 0.1202 | 0.772 | 85 | 0.6824 | 0.947 |
| 1967.BS.Esp.MiguelAngelAsturi | 4237 | 0.0123 | 0.745 | 3412 | 0.0780 | 0.614 | 804 | 0.4216 | 0.845 |
| 1971.BS.Esp.PabloNeruda.txt | 2326 | 0.0181 | 0.773 | 1789 | 0.0900 | 0.669 | 468 | 0.4466 | 0.859 |
| 1973.Bando Nro 5.txt | 4601 | 0.0130 | 0.717 | 3711 | 0.0738 | 0.599 | 801 | 0.4569 | 0.860 |
| 1973.Salvador Allende.txt | 3809 | 0.0139 | 0.748 | 2938 | 0.0841 | 0.622 | 700 | 0.4486 | 0.868 |
| 1976.Jorge Videla.txt | 3380 | 0.0148 | 0.753 | 2556 | 0.0876 | 0.623 | 604 | 0.4371 | 0.875 |
| 1977.BS.Esp.VicenteAleixandre | 1265 | 0.0348 | 0.775 | 1059 | 0.1020 | 0.698 | 241 | 0.5685 | 0.917 |
| 1978.Juan Carlos I.txt | 5507 | 0.0096 | 0.737 | 4345 | 0.0769 | 0.584 | 973 | 0.4224 | 0.848 |
| 1982.BS.Esp.GabrielGarciaMar | 2738 | 0.0175 | 0.751 | 2188 | 0.0905 | 0.641 | 522 | 0.4808 | 0.876 |
| 1982.Leopoldo Galtieri.txt | 694 | 0.0634 | 0.778 | 634 | 0.1025 | 0.754 | 119 | 0.6387 | 0.934 |
| 1990.BS.Esp.OctavioPaz.txt | 3345 | 0.0158 | 0.740 | 2619 | 0.0909 | 0.625 | 613 | 0.4633 | 0.878 |
| 2004.Pilar Manjón.txt | 1149 | 0.0409 | 0.768 | 950 | 0.1095 | 0.702 | 209 | 0.5646 | 0.917 |
| 2008.J. L. Rodriguez Zapatero.t: | 2549 | 0.0177 | 0.767 | 2005 | 0.0853 | 0.647 | 449 | 0.4543 | 0.886 |
| 2008.Julio Cobos.txt | 1443 | 0.0340 | 0.755 | 1210 | 0.0901 | 0.688 | 280 | 0.4929 | 0.907 |
| 2010.BS.Esp.MarioVargasLlosa | 2179 | 0.0225 | 0.752 | 1756 | 0.0928 | 0.654 | 424 | 0.4811 | 0.888 |
| 2010.Raúl Castro.txt | 1415 | 0.0339 | 0.765 | 1158 | 0.1071 | 0.695 | 260 | 0.5577 | 0.912 |
| 2010.Sebastian Pinera Echeniq | 2203 | 0.0213 | 0.757 | 1718 | 0.0902 | 0.648 | 432 | 0.4005 | 0.890 |
| 1868.CarlosMCespedes.txt | 8081 | 0.0063 | 0.736 | 6361 | 0.0629 | 0.582 | 1457 | 0.4056 | 0.836 |
| 1819.Simon Bolivar.txt | 63674 | 0.0011 | 0.696 | 50374 | 0.0340 | 0.488 | 11502 | 0.2286 | 0.751 |
| 1912.Emiliano Zapata.txt | 14493 | 0.0041 | 0.715 | 11946 | 0.0483 | 0.550 | 2590 | 0.3610 | 0.811 |
| 1917.Emiliano Zapata.txt | 9001 | 0.0059 | 0.732 | 6967 | 0.0693 | 0.563 | 1619 | 0.4033 | 0.826 |
| 1918.Emiliano Zapata.txt | 8025 | 0.0070 | 0.724 | 6515 | 0.0597 | 0.584 | 1438 | 0.4124 | 0.830 |
| 1933.JAntonioPrimoDeRivera.t | 16896 | 0.0039 | 0.692 | 13498 | 0.0528 | 0.528 | 3190 | 0.3047 | 0.803 |
| 1946.Jorge Eliecer Gaitan.txt | 18953 | 0.0034 | 0.704 | 14620 | 0.0561 | 0.531 | 3544 | 0.2782 | 0.811 |
| 1952.Eva Perón.txt | 5672 | 0.0100 | 0.721 | 4383 | 0.0719 | 0.588 | 1124 | 0.3060 | 0.839 |
| 1959.Fidel Castro.txt | 15237 | 0.0050 | 0.702 | 11936 | 0.0567 | 0.540 | 2892 | 0.2950 | 0.810 |
| 1964.Ernesto Che Guevara.txt | 40987 | 0.0020 | 0.672 | 32534 | 0.0432 | 0.497 | 7172 | 0.2665 | 0.779 |
| 1967.Ernesto Che Guevara.txt | 33029 | 0.0023 | 0.681 | 26129 | 0.0460 | 0.508 | 5870 | 0.2891 | 0.788 |
| 1967.Fidel Castro.txt | 30129 | 0.0023 | 0.693 | 23837 | 0.0424 | 0.516 | 5519 | 0.2232 | 0.788 |
| 1967.NL.Esp.MiguelAngelAsturi | 26424 | 0.0030 | 0.674 | 21555 | 0.0463 | 0.513 | 4901 | 0.3128 | 0.787 |
| 1970.Salvador Allende.txt | 11048 | 0.0056 | 0.709 | 8734 | 0.0600 | 0.563 | 1865 | 0.3850 | 0.834 |

| | At Char Scale | | | At Fundamental Scale | | | At Word Scale | | |
|---|---|---|---|---|---|---|---|---|---|
| | *L* | *d* | *h* | *L* | *d* | *h* | *L* | *d* | *h* |
| *Message Name* | [chrs] | [0-1] | [0-1] | [F.S.] | [0-1] | [0-1] | [w] | [0-1] | [0-1] |
| 1971.Pablo Neruda.txt | 19893 | 0.0031 | 0.704 | 15712 | 0.0538 | 0.532 | 3683 | 0.3503 | 0.806 |
| 1972.Salvador Allende.txt | 42804 | 0.0015 | 0.706 | 33483 | 0.0429 | 0.501 | 7417 | 0.2694 | 0.778 |
| 1973.Augusto Pinochet.txt | 23950 | 0.0025 | 0.714 | 18621 | 0.0491 | 0.527 | 4193 | 0.3146 | 0.797 |
| 1977.NL.Esp.VicenteAleixandre | 12379 | 0.0056 | 0.695 | 10068 | 0.0609 | 0.552 | 2379 | 0.3611 | 0.818 |
| 1979.Adolfo Suarez.txt | 79333 | 0.0009 | 0.679 | 61701 | 0.0349 | 0.476 | 13201 | 0.2120 | 0.751 |
| 1979.Fidel Castro.txt | 74583 | 0.0011 | 0.669 | 59493 | 0.0320 | 0.476 | 12838 | 0.2078 | 0.743 |
| 1981.Adolfo Suarez.txt | 7346 | 0.0065 | 0.751 | 5531 | 0.0674 | 0.574 | 1348 | 0.3116 | 0.842 |
| 1981.Roberto Eduardo Viola.txt | 23067 | 0.0029 | 0.698 | 18209 | 0.0499 | 0.525 | 3823 | 0.3369 | 0.799 |
| 1982.Gabriel Garcia Marquez.t | 11419 | 0.0061 | 0.693 | 9358 | 0.0550 | 0.555 | 2095 | 0.4086 | 0.831 |
| 1982.Felipe González.txt | 38382 | 0.0019 | 0.681 | 30636 | 0.0416 | 0.499 | 6592 | 0.2758 | 0.782 |
| 1983.Raul Alfonsin.txt | 18599 | 0.0034 | 0.704 | 14833 | 0.0501 | 0.538 | 3309 | 0.2950 | 0.805 |
| 1987.Camilo Jose Cela.txt | 8301 | 0.0076 | 0.714 | 6554 | 0.0647 | 0.575 | 1591 | 0.3903 | 0.830 |
| 1989.Carlos Saul Menem.txt | 6450 | 0.0085 | 0.732 | 4966 | 0.0763 | 0.580 | 1199 | 0.3369 | 0.845 |
| 1989.NL.Esp.CamiloJoseCela.tx | 33979 | 0.0022 | 0.676 | 27214 | 0.0419 | 0.509 | 6293 | 0.2867 | 0.777 |
| 1990.NL.Esp.OctavioPaz.txt | 25831 | 0.0029 | 0.685 | 20968 | 0.0460 | 0.518 | 4836 | 0.3002 | 0.788 |
| 1992.Rafael Caldera.txt | 14167 | 0.0045 | 0.700 | 11091 | 0.0572 | 0.539 | 2504 | 0.3323 | 0.810 |
| 1996.Jose Maria Aznar.txt | 29982 | 0.0022 | 0.699 | 24043 | 0.0403 | 0.522 | 5071 | 0.2727 | 0.782 |
| 1999.Hugo Chavez.txt | 66784 | 0.0013 | 0.667 | 53936 | 0.0329 | 0.482 | 12768 | 0.1912 | 0.760 |
| 2000.Vicente Fox.txt | 42804 | 0.0015 | 0.706 | 33483 | 0.0429 | 0.501 | 7417 | 0.2694 | 0.778 |
| 2001.Fernando de la Rua.txt | 6342 | 0.0090 | 0.723 | 4926 | 0.0717 | 0.588 | 1129 | 0.3862 | 0.853 |
| 2005.Daniel Ortega.txt | 40751 | 0.0019 | 0.689 | 32819 | 0.0260 | 0.530 | 7651 | 0.1979 | 0.778 |
| 2006.Alvaro Uribe.txt | 26323 | 0.0026 | 0.695 | 21129 | 0.0475 | 0.522 | 4555 | 0.3407 | 0.776 |
| 2006.Evo Morales.txt | 18755 | 0.0042 | 0.680 | 14748 | 0.0500 | 0.527 | 3393 | 0.2626 | 0.812 |
| 2006.Gaston Acurio.txt | 24311 | 0.0029 | 0.689 | 19835 | 0.0437 | 0.525 | 4360 | 0.2927 | 0.803 |
| 2006.Hugo Chavez.txt | 18043 | 0.0040 | 0.697 | 14287 | 0.0538 | 0.536 | 3353 | 0.2827 | 0.808 |
| 2007.Cristina Kirchner.txt | 27524 | 0.0027 | 0.692 | 21156 | 0.0487 | 0.506 | 5008 | 0.2452 | 0.795 |
| 2007.Daniel Ortega.txt | 18653 | 0.0039 | 0.687 | 14823 | 0.0487 | 0.533 | 3373 | 0.2541 | 0.805 |
| 2010.NL.Esp.MarioVargasLlosa | 37797 | 0.0021 | 0.677 | 30184 | 0.0433 | 0.507 | 7034 | 0.3149 | 0.763 |
| CamiloJoseCela.LaColmena.Nc | 8041 | 0.0077 | 0.710 | 6548 | 0.0640 | 0.573 | 1623 | 0.3672 | 0.829 |
| GabrielGMarquez.DicursoCarta | 7397 | 0.0088 | 0.706 | 5891 | 0.0689 | 0.576 | 1443 | 0.4012 | 0.844 |
| GabrielGMarquez.MejorOficioD | 16483 | 0.0036 | 0.712 | 12920 | 0.0566 | 0.537 | 2949 | 0.3591 | 0.808 |
| MarioVargasLlosa.DiscursoBue | 10772 | 0.0061 | 0.700 | 8661 | 0.0637 | 0.555 | 1986 | 0.3912 | 0.820 |
| OctavioPaz.DiscursoZacatecas. | 11767 | 0.0048 | 0.718 | 0 | 0.0576 | 0.551 | 2238 | 0.3177 | 0.810 |

ProgramingCodeProperties.htm

| | | | | | | | | | |
|---|---|---|---|---|---|---|---|---|---|
| BoolFunctWithMultiplexerLogic.C.t | 3369 | 0.0258 | 0.7191 | 2751 | 0.0712 | 0.634 | 1112 | 0.1457 | 0.794 |
| ChainedScatterTable.CSharp.txt | 793 | 0.0668 | 0.785 | 605 | 0.1306 | 0.727 | 201 | 0.2289 | 0.890 |
| CopyFolderNContent.CSharp.txt | 986 | 0.0517 | 0.811 | 687 | 0.1223 | 0.743 | 203 | 0.2463 | 0.910 |
| ExtendedEuclidean.C.txt | 200 | 0.1650 | 0.704 | 187 | 0.1872 | 0.679 | 88 | 0.2841 | 0.904 |
| Factorial.CSharp.txt | 138 | 0.2174 | 0.837 | 111 | 0.2793 | 0.802 | 40 | 0.5500 | 0.961 |
| FibonacciNumbers.CSharp.txt | 229 | 0.1485 | 0.809 | 174 | 0.2299 | 0.751 | 64 | 0.4375 | 0.924 |
| GameOfLife.C.txt | 729 | 0.0782 | 0.763 | 499 | 0.1403 | 0.647 | 247 | 0.1862 | 0.893 |
| HanoiTowers.Java.txt | 2055 | 0.0345 | 0.761 | 1526 | 0.0858 | 0.683 | 512 | 0.1816 | 0.846 |
| HeapSort.CSharp.txt | 768 | 0.0625 | 0.759 | 524 | 0.1221 | 0.658 | 261 | 0.1801 | 0.901 |
| HeapSort.Java.txt | 1110 | 0.0514 | 0.746 | 808 | 0.1002 | 0.663 | 340 | 0.1765 | 0.854 |
| InsertAfterBefore.CSharp.txt | 611 | 0.0655 | 0.813 | 393 | 0.1552 | 0.731 | 141 | 0.2624 | 0.935 |
| IsPrime.C.txt | 572 | 0.1049 | 0.766 | 431 | 0.1740 | 0.734 | 162 | 0.3519 | 0.905 |
| Levenberg.MathLab.txt | 1728 | 0.0324 | 0.733 | 1214 | 0.1112 | 0.625 | 579 | 0.1658 | 0.826 |
| MathLab.Fr.MathLab.txt | 5579 | 0.0133 | 0.704 | 4098 | 0.0639 | 0.591 | 1723 | 0.1207 | 0.790 |
| MatrixLUDecomp.CSharp.txt | 1183 | 0.0482 | 0.730 | 884 | 0.0894 | 0.633 | 420 | 0.1262 | 0.858 |
| MatrixLUDecomp.Phyton.txt | 2084 | 0.0331 | 0.739 | 1610 | 0.0981 | 0.604 | 702 | 0.1624 | 0.783 |
| MetaWords.FormsAnsClasses.CSha | 6069 | 0.0129 | 0.774 | 4863 | 0.0506 | 0.651 | 1341 | 0.1081 | 0.826 |
| ModularInverse.C.txt | 220 | 0.1727 | 0.693 | 218 | 0.1743 | 0.689 | 95 | 0.3368 | 0.906 |
| PartDifEqtnsLaplaceEq.MathLab.txt | 2113 | 0.0289 | 0.685 | 1574 | 0.0807 | 0.585 | 843 | 0.1163 | 0.780 |
| PartDifEqtnsWaveEqtn.MathLab.txt | 670 | 0.0821 | 0.732 | 569 | 0.1318 | 0.682 | 249 | 0.2731 | 0.854 |
| PermutationAlgorithm.Csharp.txt | 2499 | 0.0256 | 0.725 | 1591 | 0.0886 | 0.614 | 825 | 0.1079 | 0.839 |
| PermutationAlgorithm.Java.txt | 5913 | 0.0108 | 0.766 | 3024 | 0.0612 | 0.582 | 1305 | 0.0743 | 0.776 |
| Polinom.CSharp.txt | 396 | 0.1111 | 0.812 | 226 | 0.1947 | 0.755 | 92 | 0.3696 | 0.917 |
| QuadraticPrograming.CSharp.txt | 2078 | 0.0322 | 0.785 | 934 | 0.1167 | 0.645 | 485 | 0.1464 | 0.847 |
| QuickSort.CSharp.txt | 1204 | 0.0507 | 0.744 | 759 | 0.1120 | 0.648 | 376 | 0.1516 | 0.898 |
| SnakeGame.C.txt | 4833 | 0.0166 | 0.734 | 4415 | 0.0392 | 0.661 | 1545 | 0.1003 | 0.804 |
| Sumation.CSharp.txt | 360 | 0.1000 | 0.849 | 188 | 0.1915 | 0.746 | 71 | 0.3521 | 0.895 |
| BlowfishEncryption.C.txt | 22732 | 0.0040 | 0.741 | 17075 | 0.0362 | 0.628 | 4808 | 0.2552 | 0.674 |
| FiniteElements.MathLab.txt | 8128 | 0.0095 | 0.700 | 6703 | 0.0513 | 0.574 | 2802 | 0.1056 | 0.732 |
| FTPFunctions.CSharp.txt | 34172 | 0.0026 | 0.758 | 21721 | 0.0335 | 0.571 | 7410 | 0.0421 | 0.714 |
| MathLab.programa2.MathLab.txt | 25429 | 0.0040 | 0.652 | 16954 | 0.0363 | 0.487 | 9346 | 0.0289 | 0.686 |
| MathLab.Taller.MathLab.txt | 9531 | 0.0069 | 0.761 | 4405 | 0.0661 | 0.560 | 2166 | 0.0559 | 0.741 |
| MatrixFuncts.CSharp.txt | 16318 | 0.0053 | 0.691 | 11930 | 0.0370 | 0.532 | 5801 | 0.0336 | 0.742 |
| MetaWordsMainForm.CSharp.txt | 186597 | 0.0006 | 0.748 | 124312 | 0.0229 | 0.515 | 41392 | 0.0272 | 0.648 |
| NetPlex.Classes.CSharp.txt | 86696 | 0.0010 | 0.777 | 53863 | 0.0322 | 0.535 | 19976 | 0.0327 | 0.680 |
| NetPlex.Forms.CSharp.txt | 347738 | 0.0003 | 0.774 | 226256 | 0.0209 | 0.502 | 69994 | 0.0212 | 0.637 |
| NetPlexMainForm.CSharp.txt | 191198 | 0.0005 | 0.763 | 139473 | 0.0219 | 0.509 | 40258 | 0.0305 | 0.633 |
| Sociodynamica.Forms.txt | 9994 | 0.0080 | 0.762 | 6872 | 0.0534 | 0.623 | 2498 | 0.1201 | 0.759 |
| Sociodynamica.Module1.txt | 44370 | 0.0018 | 0.742 | 22334 | 0.0365 | 0.529 | 10263 | 0.0284 | 0.665 |
| Sociodynamica.Module2.txt | 27903 | 0.0032 | 0.738 | 16273 | 0.0482 | 0.523 | 7932 | 0.0538 | 0.706 |
| Sociodynamica.Module3.txt | 11890 | 0.0067 | 0.744 | 7673 | 0.0508 | 0.574 | 3599 | 0.0622 | 0.763 |
| ViscomSoft.ScannerActivex.CSharp. | 31157 | 0.0031 | 0.748 | 23600 | 0.0363 | 0.572 | 6483 | 0.0956 | 0.684 |
| WebSite.Inmogal.php.txt | 75224 | 0.0014 | 0.705 | 42260 | 0.0363 | 0.493 | 19279 | 0.0339 | 0.631 |
| WebSite.RistEuropa.Html.txt | 47289 | 0.0021 | 0.692 | 25804 | 0.0350 | 0.494 | 11715 | 0.0430 | 0.595 |
| WebSite.TiempoReal.Html.txt | 33799 | 0.0026 | 0.709 | 17610 | 0.0474 | 0.506 | 7509 | 0.0752 | 0.587 |

MIDIMusicProperties.htm

# MIDI Music properties At Fundamental Scale

| | | |
|---|---|---|
| *N = Length* | *[F.S.] = [Fundamental Symbols]* | |
| *d = Specific Diversity* | *[chrs] = [characters]* | |
| *h = entropy* | *[w] = [words]* | *[0-1] = between 0 and 1* |

| | | | At Fundamental Scale | | |
|---|---|---|---|---|---|
| | | | *N* | *d* | *h* |
| *Period/Style* | Pieces | *Composers* | [F.S.] | [0-1] | [0-1] |
| Total | 438 | >71 | | | |
| Medieval | | 12 | 182839 | 0.0143 | 0.6281 |
| Reinainssance | | 10 | 308629 | 0.0215 | 0.5952 |
| Baroque | | 8 | 1396651 | 0.0222 | 0.5386 |
| Classical | | 7 | 2409305 | 0.0222 | 0.5343 |
| Romantic | | 13 | 4809201 | 0.0202 | 0.5182 |
| Impressionistic | | 4 | 1363204 | 0.0244 | 0.5139 |
| 20th Century | | 8 | 1455986 | 0.0243 | 0.5220 |
| Chinese | | Several | 474214 | 0.0304 | 0.5414 |
| Hindu-Raga | | Several | 109055 | 0.0409 | 0.6220 |
| Movie Themes | | Several | 400105 | 0.0349 | 0.5781 |
| Rock | | 5 | 619413 | 0.0287 | 0.4968 |
| Venezuelan | | >20 | 894249 | 0.0273 | 0.4549 |